\documentclass[prd,12pt,nofootinbib,preprint]{revtex4}

\usepackage[T1]{fontenc}
\usepackage{amsmath,amssymb}
\usepackage{relsize}
\usepackage{epsfig}
\usepackage{url}
\usepackage{subfigure}
\usepackage{graphicx}
\usepackage{grffile}
\usepackage[usenames,dvipsnames]{color}
\usepackage{slashed}
\usepackage{array}
\usepackage{multirow}
\usepackage[colorlinks,citecolor=blue]{hyperref}
\usepackage{color}
\usepackage{cancel}
\usepackage{gensymb}
\usepackage{soul}

\def \met{\slashed{E}_T }

\def\a {\alpha}
\def\b {\beta}

\def\D {\Delta}
\def\s {\sigma}
\def\k {\kappa}

\def\l {\lambda}

\def\bar {\overline}

\def\be {\begin{equation}}
\def\ee {\end{equation}}
\def\beq {\begin{equation}}
\def\eeq {\end{equation}}
\def\bea {\begin{eqnarray}}
\def\eea {\end{eqnarray}}

\newcommand{\besub}{\begin{subequations}}
\newcommand{\eesub}{\end{subequations}}

\def\beq{\begin{equation}}
\def\eeq{\end{equation}}
\def\barr{\begin{array}}
\def\earr{\end{array}}

\begin{document}
%%%%%%%%%%%%%%%%%%%%%%%%%%%%%%%%%%%%%%%%%%%%%%%%%%%%%%%%%%%%%%%%%%%%%%%%%%%%%%%%%
\title{Muon $g-2$ in a Type-X 2HDM assisted by inert scalars: probing at the LHC}

\author{Nabarun Chakrabarty}
\email{nabarunc@iitk.res.in}
\affiliation{Department of Physics, Indian Institute of Technology Kanpur, Kanpur-208016, Uttar Pradesh, India}

\begin{abstract} 
A scenario augmenting the well known Type-X Two-Higgs doublet model (2HDM) with an additional inert doublet is proposed. The Type-X 2HDM is known to offer a solution to the muon $g-2$ anomaly for a light pseudoscalar. We show that  
the proposed framework can accomodate a heavier pseudoscalar on account of two-loop Barr-Zee (BZ) contributions to muon $g-2$ stemming from the inert doublet. We subsequently explore an interesting $\tau^+\tau^- + $ missing transverse energy signal that can be used to probe the present scenario at the 14 TeV LHC.
\end{abstract} 
%%%%%%%%%%%%%%%%
\maketitle

%%%%%%%%%%%%%%%%%%%%%%%%%%%%%%%%%%
\section{Introduction} 
%%%%%%%%%%%%%%%%%%%%%%%%%%%%%%%%%%
One of the experimental findings that continues to advocate dynamics beyond the Standard Model (SM) of particle physics is the longstanding muon $g-2$ anomaly. In a nutshell, a discrepancy has been noted between the SM prediction of the anomalous magnetic moment of the muon~\cite{Blum:2013xva,RBC:2018dos,Keshavarzi:2018mgv,Davier:2019can,Aoyama:2020ynm,Colangelo:2018mtw,Hoferichter:2019mqg,Melnikov:2003xd,Hoferichter:2018kwz,Blum:2019ugy,ParticleDataGroup:2020ssz} and its experimental measurements made at BNL~\cite{Muong-2:2006rrc} and FNAL~\cite{Muong-2:2021ojo,Muong-2:2021vma}. The combined excess reported is 
\bea
\Delta a_\mu \equiv a^{\text{exp}}_\mu - a^{\text{SM}}_\mu = 
251(59) \times 10^{-11}.
\eea  
A minimal extension of the SM long known to resolve the muon $g-2$ anomaly is the Type-X 2HDM~\cite{Deshpande:1977rw,Branco:2011iw}. The model comprises an additional $SU(2)_L$ scalar doublet over and above SM. This entails an enlarged scalar sector, i.e., two CP-even neutral scalars $h,~H$, a CP-odd neutral scalar $A$ and a charged scalar 
$H^+$.
A governing $\mathbb{Z}_2$ keeps flavour changing neutral current (FCNC) at bay by demanding that a particular fermion interacts with only one of the two doublets. This leads to several variants and one such is the Type-X 2HDM~\cite{Branco:2011iw}. One important feature of the Type-X is that quark Yukawa couplings involving the additional scalars are suppressed while the leptonic Yukawas are enhanced. It is the enhanced leptonic couplings that potentially give rise to sizeable muon $g-2$ contributions at the one-loop and two-loop levels. A resolution of the anomaly thus becomes possible for a light pseudoscalar ($M_A \lesssim$ 100 GeV) and high 
tan$\b$ (tan$\b \gtrsim$ 20)~\cite{Broggio:2014mna,Cao:2009as,Wang:2014sda,Ilisie:2015tra,Abe:2015oca,Chun:2016hzs,Cherchiglia:2016eui,Dey:2021pyn,Han:2018znu}. As a perk, by virtue of the small quark couplings, such a parameter region in the Type-X evades stringent constraints from flavour observables and direct search from the colliders~\cite{Chowdhury:2017aav}, especially the Large Hadron Collider (LHC). However, recent LHC searches for $h_{125} \to A A \to 4\tau,2\tau2\mu$ channels have disfavoured a large BR($h_{125} \to A A$)~\cite{CMS:2018qvj}. This strongly constrains the $M_A < 62.5$ GeV parameter region. Furthermore, the large 
tan$\b$ region also get restricted by lepton precision observables. A partial list of collider probes of the Type-X 2HDM is \cite{Chun:2017yob,Chun:2018vsn,Crivellin:2015hha,Iguro:2019sly,Jueid:2021avn,Chen:2021rnl,Wang:2018hnw}. 

In this study, we seek to alleviate the aforementioned shortcoming by extending the scalar sector of the Type-X 2HDM so that additional two-loop Barr-Zee (BZ) contributions
are encountered\footnote{Some studies addressing the same problem using additional fermionic content and above the Type-X 2HDM are \cite{Frank:2020smf,Chun:2020uzw}}. The simplest multiplets leading to additional BZ diagrams are $SU(2)_L$ singlets with $Y \neq 0$. For instance, let us look more closely at the $Y=1$ case which is basically a singly charged scalar. This scalar can mix with the like-charge states in the 2HDM through the scalar potential and therefore can be searched at a collider via the $k^+ \to \ell^+ \nu$. Here, $k^+$ refers to the charged scalar and $\ell = e,\mu,\tau$. While this is enticing, the scope to directly look for the scalar via the invariant mass handle is obviated owing to the presence of the neutrino(s). As for an $SU(2)_L$-singlet $Y \geq 2$ scalar (that is, $Q \geq 2$), there would be no mixing of the same with the 2HDM sector that features neutral and singly charged scalars only. Such a case therefore will also not be attractive from the collider perspective.  Neutral scalars offer the possibility of having opposite-sign lepton final states at a collider. Thus, while we postulate the existence of charged scalars to induce BZ diagrams, we also aim to probe such a framework via decays of neutral scalars. And, the lowest multiplet that features both charged and neutral scalars is (\textbf{2},$\frac{1}{2}$). We have therefore augmented the Type-X 2HDM with an additional scalar $SU(2)_L$ doublet. An extra $\mathbb{Z}^\prime_2$ is imposed under which the new doublet is odd while all other fields are even. This addition can be deemed natural since the number of scalar doublets cannot be limited by the 
electroweak (EW) 
$\rho$-parameter or other fundamental constraints. While the $\mathbb{Z}_2$ is allowed to be broken by the scalar potential, 
the $\mathbb{Z}^\prime_2$ is \emph{exact} and therefore disallows mixing between the third doublet with the first two. The third doublet thus remains \emph{inert}. We refer to this scenario as (2+1)HDM and is therefore a generalisation of the popular inert doublet model (IDM). The (2+1)HDM has generated some interest in the past. Reference \cite{Moretti:2015cwa} studied in detail the constraints on this scenario from perturbative unitarity and oblique corrections. The reference \cite{Moretti:2015tva} computed the strength of the $H^\pm W^\mp Z(\gamma)$ vertex at one-loop with a particular emphasis of the contribution coming from the inert doublet. A more recent study is \cite{Merchand:2019bod}
that computed various mono-object signals that arise as predictions of the model.

In this work, we compute in detail the two-loop BZ contributions to 
$\Delta a_\mu$ stemming from the inert doublet. More precisely, we can expect sizeable contributions from such loops owing to the possibility of large scalar trilinear couplings involving the inert scalars.  
We fold-in all relevant theoretical and experimental constraints in the process including dark matter direct detection. Our aim is to see if the (2+1)HDM can expand the parameter region in the 
$M_A$ - tan$\beta$ plane that is compliant with the muon $g-2$ excess.

The (2+1)HDM also predicts an interesting collider signal. The CP-even and CP-odd components of the inert doublet can be pair produced at the LHC with following which the odd component can decay to the even component and $A$. Since, 
the $A$ in this scenario can be potentially heavier than one in the Type-X 2HDM and still comply with $\Delta a_\mu$, a $\tau^\pm$ coming from $A \to \tau^+ \tau^-$ will accordingly have a higher transverse momentum in comparison to what is seen in case of the Type-X 2HDM. Moreover, the final state will also have a modified missing-transverse energy spectrum due to the presence of inert scalars.
In all, such kinematic features of this signal makes the (2+1)HDM discernible from the Type-X 2HDM. We have performed detailed analyses of the signal and backgrounds in this work using multivariate techniques and estimated the observability at the 14 TeV LHC.

This study is segmented as follows. We introduce the (2+1)HDM scenario in 
section \ref{model} and elaborate all the constraints in section \ref{constraints}. The details of muon $g-2$ calculation can be found in section \ref{gmt}. The collider analysis is given in section \ref{collider} and we conlude in section \ref{conclusions}. Various important formulae are relegated to the Appendix.

\section{Theoretical framework}\label{model}
The 2HDM, that features two scalar doublets $\phi_1,\phi_2$ is augmented by an additional scalar doublet $\eta$ in this work. A discrete symmetry $\mathbb{Z}^\prime_2$ is introduced under which $\phi_{1,2}\to \phi_{1,2}$ while $\eta \to -\eta$. The most general scalar potential consistent with the gauge and the $\mathbb{Z}^\prime_2$ symmetry then reads:
\besub
\bea
V &=& V_2 + V_4^a + V_4^b, \label{scalar_pot} \\
\text{with} \nonumber \\
V_2 &=& - m_{11}^2 |\phi_1|^2 - m_{22}^2 |\phi_2|^2
+ m_{12}^2 (\phi_1^\dagger \phi_2 + \text{h.c.})
+ \mu^2 |\eta|^2, \\
V_4^{\{\phi_1,\phi_2\}} &=& \frac{\l_1}{2}|\phi_1|^4 + \frac{\l_2}{2}|\phi_2|^4
 + \l_3 |\phi_1|^2 |\phi_2|^2 
+ \l_4 |\phi_1^\dagger \phi_2|^2 + \frac{\l_5}{2} [(\phi_1^\dagger \phi_2)^2 + h.c.] \nonumber \\
&&
+ \l_6 [(\phi_1^\dagger \phi_1)(\phi_1^\dagger \phi_2) + h.c.]
+ \l_7 [(\phi_2^\dagger \phi_2)(\phi_1^\dagger \phi_2) + h.c.], \\
V_4^{\{\phi_1,\phi_2,\eta\}} &=& \frac{\l^\prime}{2}|\eta|^4
+ \sum_{i=1,2} \bigg\{  \nu_i |\phi_i|^2 |\eta|^2
+ \omega_i |\phi_1^\dagger \eta|^2
+ \Big[ \frac{\kappa_i}{2} (\phi_i^\dagger \eta)^2 + h.c. \Big] \bigg\} 
\nonumber \\
&&
+ \Big[ \sigma_1 |\eta|^2 \phi_1^\dagger \phi_2 + \sigma_2 \phi_1^\dagger \eta \eta^\dagger \phi_2
+ \sigma_3 \phi_1^\dagger \eta
\phi_2^\dagger \eta + h.c. \Big].
\label{eqn:pot}
\eea
\eesub
In the above, $V_2$ combines all the dimension-2 terms of the scalar potential. The dimension-4 terms involving $\phi_1,\phi_2$ alone are given by $V_4^{\{\phi_1,\phi_2\}}$. Finally, the term $V_4^{\{\phi_1,\phi_2,\eta\}}$
contains dimension-4 terms involving all the three scalar doublets. All parameters in the scalar potential are chosen real to avoid CP-violation. Following electroweak symmetry breaking (EWSB), the scalar doublets are parameterised as:
\bea
\Phi_i = \begin{pmatrix}
\phi_i^+ \\
\frac{1}{\sqrt{2}} (v_i + h_i + i z_i)
\end{pmatrix} , (i = 1,2) ,~~~ 
\eta = \begin{pmatrix}
\eta^+ \\
\frac{1}{\sqrt{2}} (\eta_R + i \eta_I)
\end{pmatrix},
\eea
where $v_{1,2}$ are the vacuum expectation values (VEVs) and tan$\beta$ = $\frac{v_2}{v_1}$. The non-inert
particle spectrum in this case is identical with the 2HDM that consists of the neutral CP-even Higgses $h, H$, a CP-odd Higgs $A$ and a charged Higgs $H^+$. The 2 $\times$ 2 mass matrices
brought into diagonal forms by the action of the mixing angles $\a$ and $\b$. Of these, the scalar $h$ is taken to be the SM-like Higgs with mass 125 GeV. Further details on the 2HDM spectrum are skipped for brevity. 

The inert scalars coming from $\eta$ do not mix with the ones from $\phi_{1,2}$ on account of the 
$\mathbb{Z}^\prime_2$. The inert mass eigenstates
are then $\eta^+,\eta_R$ and $\eta_I$ that have the following squared masses:
\besub
\bea
m^2_{\eta^+} &=& \mu^2 + \frac{1}{2}\{\nu_1 c^2_\b + \nu_2 s^2_\b\}v^2 + \sigma_1 v^2 s_\b c_\b, \\
m^2_{\eta_R} &=& \mu^2 + \frac{1}{2}\{(\nu_1 + \omega_1 + \kappa_1) c^2_\b + (\nu_2 + \omega_2 + \kappa_2) s^2_\b\}v^2 + (\sigma_1 + \sigma_2 + \sigma_3) v^2 s_\b c_\b, \\
m^2_{\eta_I} &=& \mu^2 + \frac{1}{2}\{(\nu_1 + \omega_1 - \kappa_1) c^2_\b + (\nu_2 + \omega_2 - \kappa_2) s^2_\b\}v^2 + (\sigma_1 + \sigma_2 - \sigma_3) v^2 s_\b c_\b.
\eea
\eesub 
As for the Yukawa interactions, we take the Type-X (or lepton-specific) Lagrangian where the quarks get their masses from $\phi_2$ while the all the leptons do from $\phi_1$. The Yukawa Lagrangian can be expressed as
\bea
-\mathcal{L}_Y &=& \Big[ y_u \bar{Q_L} \tilde{\phi}_2 u_R + y_d \bar{Q_L} \phi_2 d_R + y_\ell \bar{Q_L} \phi_1 \ell_R \Big] + \text{h.c.}
\eea
The lepton Yukawa interactions in terms of the physical scalars then becomes
\bea
\mathcal{L}^\text{lepton}_Y &=& \sum_{\ell=e,\mu,\tau} \frac{m_\ell}{v} \bigg(\xi_\ell^h h \bar{\ell} \ell + \xi_\ell^H H \bar{\ell} \ell - i \xi_\ell^A A \bar{\ell} \gamma_5 \ell + \Big[ \sqrt{2} \xi^A_\ell H^+ \bar{\nu_\ell} P_R \ell + \text{h.c.} \Big] \bigg).
\eea
The various $\xi_\ell$ factors are tabulated in Table \ref{tab:xi}.
\begin{table}
\centering
\begin{tabular}{ |c c c c c c c c c c| } 
\hline
 & $\xi^h_e$ & $\xi^h_\mu$ & $\xi^h_\tau$ & $\xi^H_e$ & $\xi^H_\mu$ & $\xi^H_\tau$ & $\xi^A_e$ & $\xi^A_\mu$ & $\xi^A_\tau$ \\ \hline
Lepton-specific & $-\frac{\text{sin}\a}{\text{cos}\b}$
& $-\frac{\text{sin}\a}{\text{cos}\b}$
& $-\frac{\text{sin}\a}{\text{cos}\b}$
& $\frac{\text{cos}\a}{\text{cos}\b}$
& $\frac{\text{cos}\a}{\text{cos}\b}$
& $\frac{\text{cos}\a}{\text{cos}\b}$
& tan$\beta$ & tan$\beta$ & tan$\beta$ \\ \hline
\end{tabular}
\caption{Various Yukawa scale factors for the lepton-specific case.}
\label{tab:xi}
\end{table}

\section{Constraints}\label{constraints}

\subsection{Theoretical constraints}
The scalar potential remains perturbative if $|\l| < 4\pi, |y| < \sqrt{4\pi}$ where $\l$ and $y$ respectively denote a quartic and yukawa coupling of the theory. Also, the scalar potential remains bounded from below along various directions in field space if the following conditions are satisfied~\cite{Grzadkowski:2009bt,Merchand:2019bod,Faro:2019vcd}:
\bea
\l_1 > 0,
~\l_2 > 0,
~\l^\prime > 0, 
~\l_3 + \sqrt{\l_1 \l_2} > 0, 
~\l_3 + \l_4 - |\l_5|
+ \sqrt{\l_1 \l_2} > 0, \nonumber \\ 
\nu_1 + \sqrt{\l_1 \l^\prime} > 0,
~\nu_1 + \omega_1 - |\kappa_1| + \sqrt{\l_1 \l^\prime} > 0, \nonumber \\
\nu_2 + \sqrt{\l_2 \l^\prime} > 0,
~\nu_2 + \omega_2 - |\kappa_2| + \sqrt{\l_2 \l^\prime} > 0.
\eea
These conditions are derived assuming that the various directions in field space lie in planes spanned by two field axes. A treatment involving more general directions are computationally hefty to implement and are typically less restrictive than the two-field conditions. We have therefore neglected such possibilities in this work.

The framework is also constrained by invoking unitarity of the S-matrix. Electroweak equivalence theorem dictates that the tree-level scatterings of longitudinal components of the gauge bosons can be mapped into $2\to2$ tree-level scatterings 
involving the scalars of the theory at high energies~\cite{Lee:1977eg}. Unitarity demands that the eigenvalues of the matrices formed in the bases of various $2 \to 2$ scalar scattering states must have magnitudes that are bounded from above at $8\pi$. A detailed analysis on unitarity in three-Higgs doublet models (3HDMs) was presented in \cite{Moretti:2015cwa}. In this study, we have deduced the scattering matrices for the scalar potential of Eq.(\ref{scalar_pot}), determined the eigenvalues numerically and demanded that their magnitudes do not exceed the stipulated limit of 8$\pi$.

\subsection{Higgs signal strengths}

The signal strength for the $h \to X$ channel is defined as
\bea
\mu_X = \frac{\sigma_{pp \to h}}{\sigma^{\text{exp}}_{pp \to h}} \frac{\text{BR}(h \to X)}{\text{BR}^{\text{exp}}(h \to X)}
\eea

We adhere to $\a = \b - \frac{\pi}{2}$ in this work (known as the \emph{alignment} limit in the context of 2HDMs) in which case the couplings of the fermions and gauge bosons to $h$ become equal to the corresponding SM values. In this limit, the predicted values for the Higgs signal strength in the 
$b\bar{b},\tau^+\tau^-,W^+W^-,ZZ,gg$ channels become consistent with the  measurements at ATLAS and CMS. The only channel that can still deviate in this limit is $\gamma\gamma$ where the presence of the additional charged scalars $H^+,\eta^+$ leads to new one-loop contributions in the $h \to \gamma \gamma$ amplitude. One then has
\bea
M^{\text{NP}}_{h \to \gamma \gamma} &=& \mathlarger{\mathlarger{\sum}}_{\phi^{+} = H^+,\eta^+}
\frac{\l_{h \phi^+ \phi^-}v}
{2 m^2_{\phi^+}} A_0\bigg(\frac{m^2_h}
{4 m^2_{\phi^+}}\bigg).
\eea
The total amplitude and the decay width then become
\bea
\mathcal{M}_{h \to \gamma \gamma} &=& 
\mathcal{M}^{\text{SM}}_{h \to \gamma \gamma} +
\mathcal{M}^{\text{NP}}_{h \to \gamma \gamma}, \\
\Gamma_{h \to \gamma \gamma} &=& \frac{G_F \a^2 m_h^3}{128 \sqrt{2} \pi^3} |\mathcal{M}_{h \to \gamma \gamma}|^2.\eea
where $G_F$ is the Fermi constant. The pertinent loop functions are listed below~\cite{Djouadi:2005gi,Djouadi:2005gj}.
\besub
\bea
A_0(x) &=& -\frac{1}{x^2}\big(x - f(x)\big),  \\
\text{with} ~~f(x) &=& \text{arcsin}^2(\sqrt{x}); ~~~x \leq 1 
\nonumber \\
&=& -\frac{1}{4}\Bigg[\text{log}\frac{1+\sqrt{1 - x^{-1}}}{1-\sqrt{1 - x^{-1}}} -i\pi\Bigg]^2; x > 1.
\eea
\eesub
The charged scalars do not modify the production cross section of $h$ and the modification to the total width is tiny. The signal strength in the $\gamma\gamma$ channel is then 
\bea
\mu_{\gamma\gamma} &\simeq& \frac{\Gamma_{h \to \gamma \gamma}}{\Gamma^{\text{exp}}_{h \to \gamma \gamma}}.
\eea
The latest 13 TeV results on the diphoton signal strength from the LHC read $\mu_{\gamma\gamma}=0.99^{+0.14}_{-0.14}$ (ATLAS~\cite{ATLAS:2018hxb}) and $\mu_{\gamma\gamma}=1.18^{+0.17}_{-0.14}$ (CMS~\cite{CMS:2018piu}). Upon using the standard combination of signal strengths and uncertainties, we obtain $\mu_{\gamma\gamma} = 1.06 \pm 0.1$ and impose this constraint at 2$\sigma$. 

The branching ratio of Higgs to invisible states faces upper limits from the LHC. The most recent constraint is BR$(h \to~\text{invisible}) < 18\%$ \cite{CMS:2022qva}. We have implemented this constraint in our analysis.

\subsection{Oblique parameters}

The NP corrections induced to the oblique parameters~\cite{Peskin:1991sw} in this setup are can be split into
a contribution coming from the active doublet 
($\D X_{\text{2HDM}}$) and one from the inert doublet ($\D X_{\text{IDM}}$).
\bea
\D X &=& \D X_{\text{2HDM}}
 + \D X_{\text{IDM}} 
\eea
where $X = S,T,U$. The most constraining for scalar extensions of the SM is the $T$-parameter.
The relevant expressions for $\a = \b - \frac{\pi}{2}$ are given below~\cite{Grimus:2008nb}.
\besub
\bea
\D T_{\text{2HDM}} &=& \frac{1}{16 \pi s^2_W M_W^2}\Big[F(m^2_{H^+},m^2_H)
 + F(m^2_{H^+},m^2_A) 
- F(m^2_{H},m^2_{A})\Big], \\
\D T_{\text{IDM}} &=& \frac{1}{16 \pi s^2_W M_W^2}\Big[F(m^2_{\eta^+},m^2_{\eta_R})
 + F(m^2_{\eta^+},m^2_{\eta_I}) 
- F(m^2_{\eta_R},m^2_{\eta_I})\Big]. 
\eea
\eesub
In the above,
\bea
F(x,y) &=&  \frac{x+y}{2} - \frac{xy}{x-y}~{\rm ln} \bigg(\frac{x}{y}\bigg)~~~ {\rm for} ~~~x \neq y \,, \nonumber \\
&=& 0~~~ {\rm for} ~~~ x = y.
\eea
The most updated bound reads~\cite{ParticleDataGroup:2020ssz}
\bea
\Delta T = 0.07 \pm 0.12.
\eea
We have imposed the stated bound have been at 
2$\sigma$ in our analysis.

\subsection{Dark matter}

The presence of the inert doublet protected by a discrete symmetry makes its neutral component ($\eta_R$ or $\eta_I$) a DM candidate.
The PLANCK collaboration quotes the following as the latest measured value of the DM relic density~\cite{Planck:2018vyg}:
\bea
\Omega_{\text{Planck}} h^2 &=& 0.120 \pm 0.001
\eea
The other important DM constraint comes from the search of DM-nucleon scattering cross sections by different terrestrial experiments such as XENON-1T~\cite{XENON:2017vdw,XENON:2018voc} and PANDA-X~\cite{PandaX-II:2016vec,PandaX-II:2017hlx}.  
The non-observation of such scatterings has led to upper limits on the DM-nucleon cross section with the most stringent bound for $m_{\text{DM}} < $ 1 TeV coming from XENON-1T.

The model is implemented to the publicly available tool \texttt{micrOMEGAs}~\cite{Belanger:2014vza} in order to compute the relic density and the spin-independent direct detection (SI-DD) cross section. The computed relic density is stipulated to be underabundant in this work as presence of other DM candidates (not accounted for in this work) is assumed. Folding-in a 10$\%$ experimental error in the measured central value, we demand
\bea
\Omega h^2 < 0.12 \pm 2\times 0.012
\eea
As for direct detection, we compute the SI-DD cross section $\sigma_{\text{SI}}$ using \texttt{micrOMEGAs}. We subsequently determine the effective cross section $\sigma^{\text{eff}}_{\text{SI}} = \Big(\frac{\Omega h^2}{\Omega_{\text{Planck}}h^2}\Big)\sigma_{\text{SI}}$. This ratio takes care of the fact that the present model acconts for only a part of the observed relic abundance. We add that the SI-DD scatterings in this scenario proceeds via t-channel diagrams involving $h$ and $H$.

\section{The muon $g-2$ amplitude and its numerical prediction}\label{gmt}

\begin{figure}
\centering
\subfigure[]{
\includegraphics[scale=0.37]{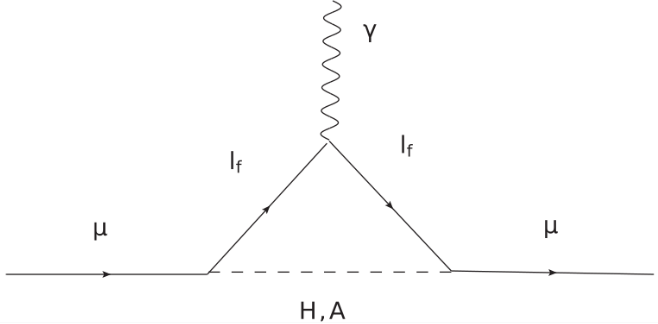}}
\subfigure[]{
\includegraphics[scale=0.34]{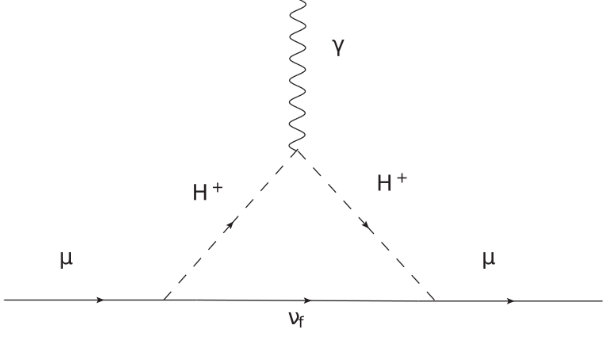}}
\caption{One loop contributions to $\Delta a_\mu$ from (a) $H,A$ and (b) $H^+$.}
\label{f:delmu_oneloop}
\end{figure}

We present an elaborate computation of $\Delta a_\mu$ in this section. The electromagnetic interaction of a lepton is given by
\bea
\bar{\ell}(p_2) \Gamma^\mu \ell(p_1) &=& \bar{\ell}(p_2) \Big[\gamma^\mu F_1(q^2) + 
\frac{i \sigma^{\mu\nu} q_\nu F_2(q^2)}{2 M_\ell} \Big] \ell(p_1)
\eea
with $F_{1,2}(q^2)$ as some form factors. The lepton anomalous magnetic dipole moment is then defined as 
\bea
a_\ell &=& F_2(0).
\eea
While giving the expressions for the various 
$\Delta a_\mu$ contributions in the (2+1)HDM, we denote the loop order in the superscript and the particle circulating in the loop in the subscript. The one-loop amplitudes driven by $H,A,H^+$ in the alignment limit are expressed below: 
\besub
\bea
{\Delta a_\mu}_{(H)}^{(1\text{loop})} &=& \frac{M_\mu^2}{8 \pi^2 v^2}  \left(\frac{M_\mu^2}{M_H^2}\right) \big(\xi_\mu^{H} \big)^2~ \int_{0}^{1} dx \frac{x^2(2-x)}{\left(\frac{M_\mu^2}{M_H^2}\right) x^2 -x + 1}, \\
{\Delta a_\mu}_{(A)}^{(1\text{loop})} &=& -\frac{M_\mu^2}{8 \pi^2 v^2}  \left(\frac{M_\mu^2}{M_A^2}\right) \big(\xi_\mu^{A} \big)^2~ \int_{0}^{1} dx \frac{x^3}{\left(\frac{M_\mu^2}{M_A^2}\right) x^2 -x + 1}, \\ 
{\Delta a_\mu}_{(H^+)}^{(1\text{loop})} &=& \frac{M_\mu^2}{8 \pi^2 v^2} \left(\frac{M_\mu^2}{M_{H^+}^2}\right)\big(\xi_\mu^A \big)^2~\int_{0}^{1} dx \frac{x^2(1-x)}{\left(\frac{M_\mu^2}{M_{H^+}^2}\right)x (1-x)-x}.
\eea
\eesub
One notes that ${\Delta a_\mu}_{(H^+)}^{(1l)} < 0$. The corresponding diagrams are shown in Fig.\ref{f:delmu_oneloop}.

\begin{figure}
\centering
\subfigure[]{
\includegraphics[scale=0.41]{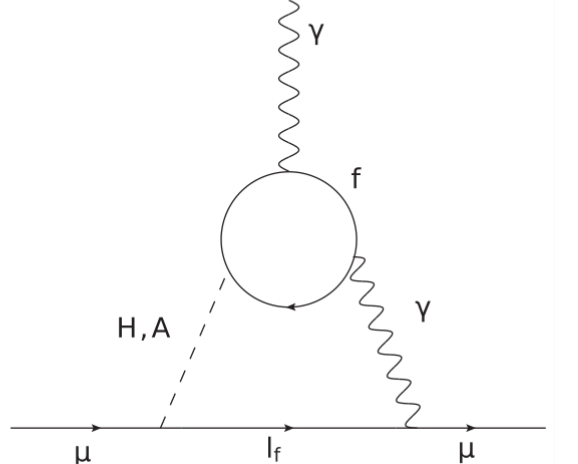}}
\subfigure[]{
\includegraphics[scale=0.13]{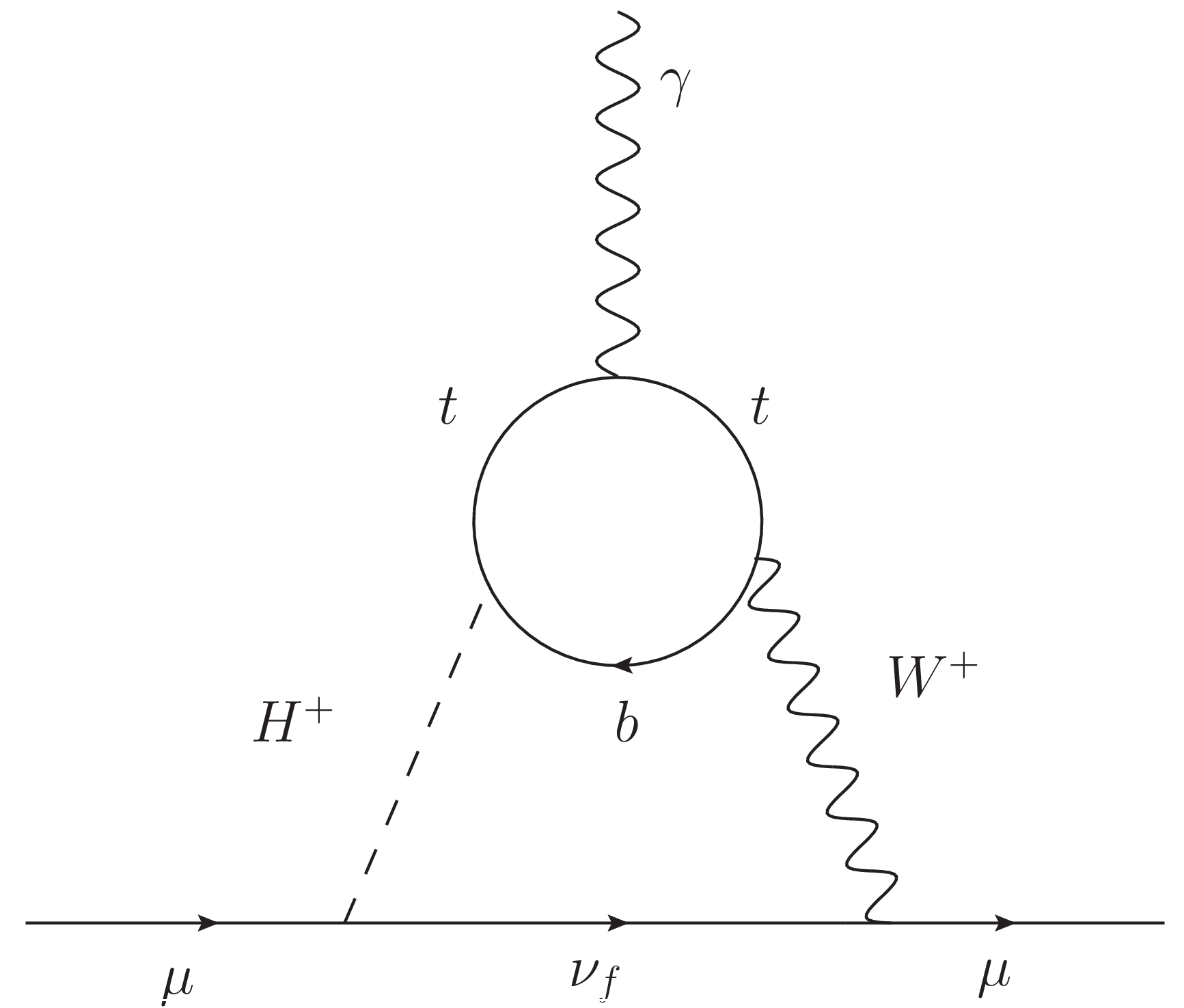}}
\caption{Two loop contributions to $\Delta a_\mu$ from the fermions through (a) an effective $\phi \gamma \gamma$ vertex and (b) an effective $H^+ W^- \gamma$ vertex.}
\label{f:delmu_twoloop_f}
\end{figure}

The two-loop BZ contributions arise upon embedding $h\gamma\gamma,H\gamma\gamma,A\gamma\gamma$ and $H^+ W^- \gamma$ form factors that themselves arise at one-loop, in a one-loop amplitude. The resulting topology is thus two-loop. We firstly list out the Feynman graphs that feature fermions in the one-loop form factors in Fig.\ref{f:delmu_twoloop_f}.

\begin{figure}
\centering
\subfigure[]{
\includegraphics[scale=0.32]{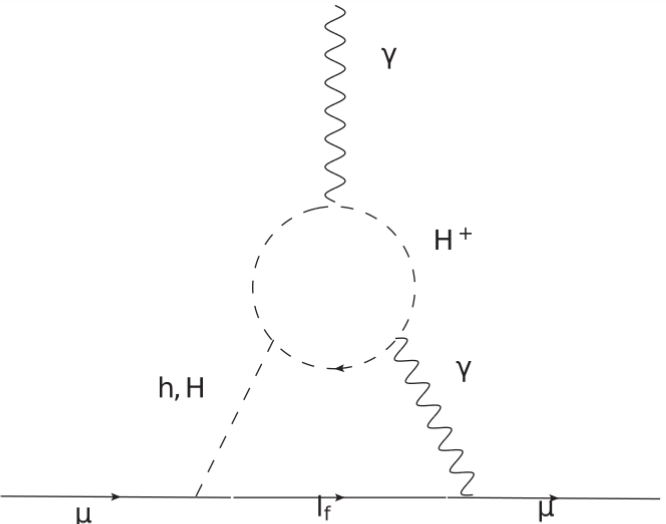}}
\subfigure[]{
\includegraphics[scale=0.32]{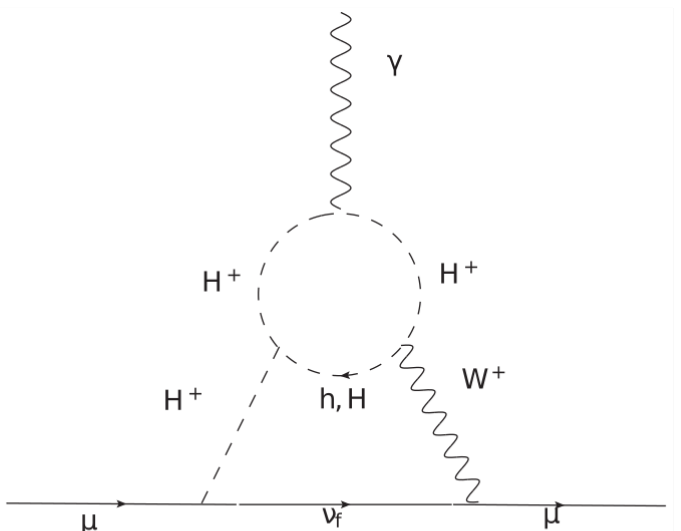}}
\caption{Two loop contributions to $\Delta a_\mu$ from the 2HDM scalars through (a) an effective $\phi \gamma \gamma$ vertex and (b) an effective $H^+ W^- \gamma$ vertex.}
\label{f:delmu_twoloop_2HDM}
\end{figure}

\besub
\bea
{\Delta a_\mu}_{\{f,H \gamma\gamma\}}^{(2\text{loop})} &=& \sum_{f} \frac{\alpha_{\text{em}} M_\mu^2}{4 \pi^3 v^2}~ N_C^f Q_f^2 \xi_f^{H} \xi_\mu^{H} \mathcal{F}^{(1)}\left(\frac{M_f^2}{M_H^2}\right), \\
{\Delta a_\mu}_{\{f,A \gamma\gamma\}}^{(2\text{loop})} &=& \sum_{f} \frac{\alpha_{\text{em}} M_\mu^2}{4 \pi^3 v^2}~ N_C^f Q_f^2 \xi_f^A \xi_\mu^A \mathcal{\tilde{F}}^{(1)}\left(\frac{M_f^2}{M_A^2}\right), \\
{\Delta a_\mu}_{\{f,~H^+ W^- \gamma\}}^{(2\text{loop})} &=& \frac{\alpha_{\text{em}} M_\mu^2 N_t |V_{tb}|^2}{32 \pi^3 s_w^2 v^2 (M_{H^+}^2 - M_W^2)} \int_{0}^{1} dx \left[Q_t x + Q_b(1-x)\right] \nonumber \\
&&
\times \left[\xi^A_d \xi^A_\mu M_b^2 x(1-x) + \xi^A_u \xi^A_\mu M_t^2 x(1+x)\right] \nonumber \\
&& \times \left[\mathcal{G}\left(\frac{M_t^2}{M_{H^+}^2},\frac{M_b^2}{M_{H^+}^2},x\right) - \mathcal{G}\left(\frac{M_t^2}{M_W^2},\frac{M_b^2}{M_W^2},x\right)\right].
\eea
\eesub
Here, $N_C^f$ = 1(3) for leptons (quarks).
Further, $\alpha_{\text{em}}$ denotes the fine structure constant and $Q_t = 2/3,~Q_b = -1/3$. Next to come are the two-loop amplitudes induced by the 2HDM scalars running in the loops as shown in Fig.\ref{f:delmu_twoloop_2HDM}. The corresponding amplitudes for $\a = \b - \frac{\pi}{2}$ are expressed below.

\begin{figure}
\centering
\subfigure[]{
\includegraphics[scale=0.32]{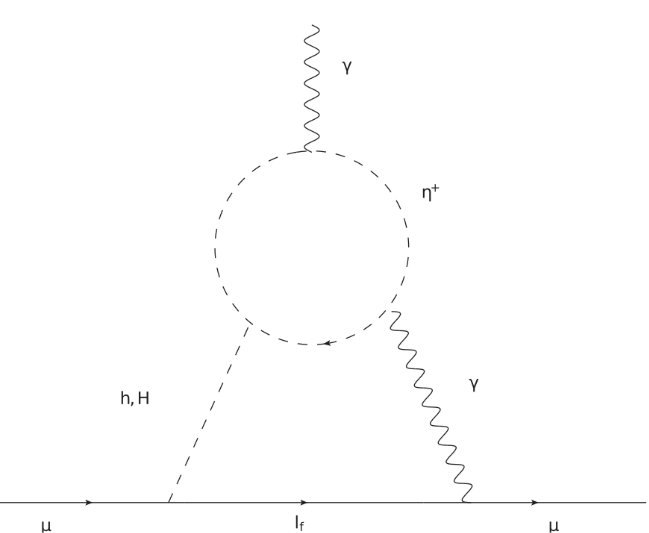}}
\subfigure[]{
\includegraphics[scale=0.32]{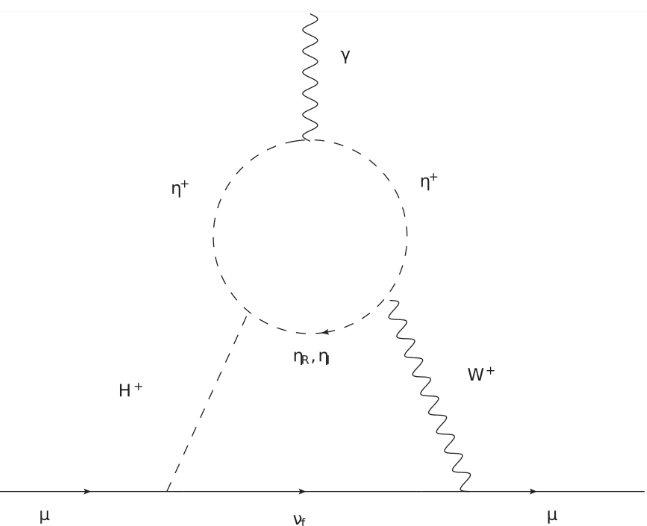}}
\caption{Two loop contributions to $\Delta a_\mu$ from the inert scalars through (a) an effective $\phi \gamma \gamma$ vertex and (b) an effective $H^+ W^- \gamma$ vertex.}
\label{f:delmu_twoloop_inert}
\end{figure}

\begin{figure}
\centering
\subfigure[]{
\includegraphics[scale=0.47]{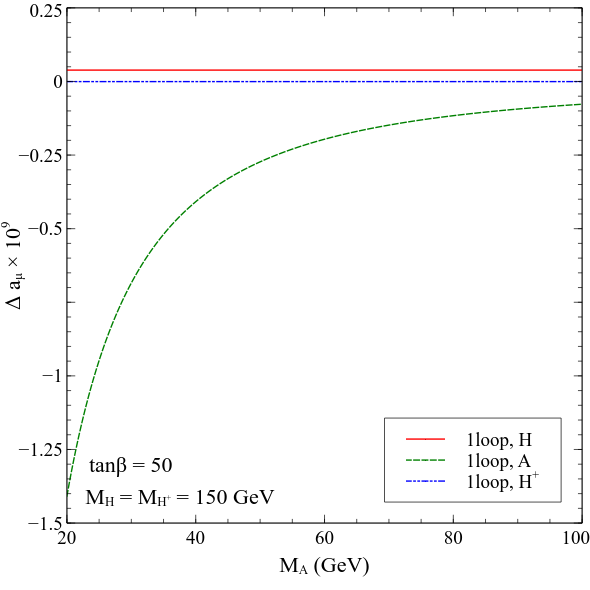}}~~~~
\subfigure[]{
\includegraphics[scale=0.47]{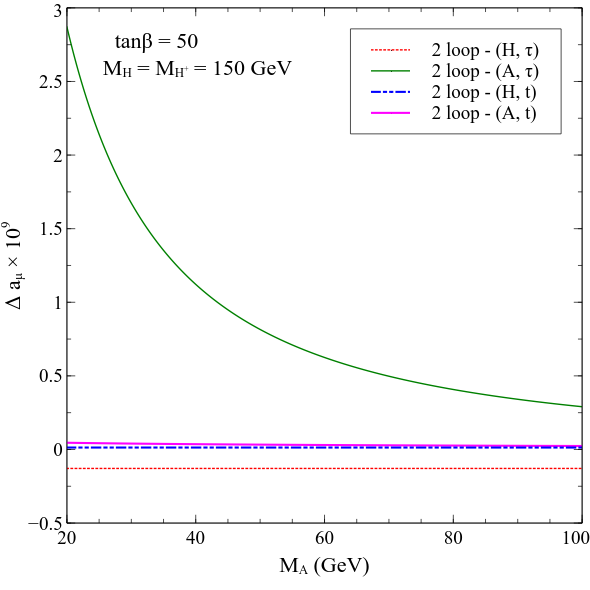}}
\caption{One-loop and two-loop fermion-mediated Barr-Zee contributions to $\Delta a_\mu$. The colour coding is explained in the legends.}
\label{f:contri1}
\end{figure}

\besub
\bea
{\Delta a_\mu}_{\{H^+,~\phi\gamma\gamma\}}^{(2\text{loop})} &=&  \sum_{\phi = h, H} \frac{\alpha_{\text{em}} M_\mu^2}{8 \pi^3 M_{\phi}^2}~ \xi_\mu^{\phi}~ \lambda_{\phi H^+ H^-}\mathcal{F}^{(2)}\left(\frac{M_{H^+}^2}{M_{\phi}^2}\right), \\
{\Delta a_\mu}_{\{H,~H^+ W^-\gamma\}}^{(2\text{loop})} &=& \frac{\alpha_{\text{em}} M_\mu^2 }{64 \pi^3 s_w^2 (M_{H^+}^2 - M_W^2)}  \xi^{H}_\mu ~\lambda_{H H^+ H^-} \int_{0}^{1} dx~x^2 (x-1) \nonumber \\
&&\times \left[\mathcal{G}\left(1,\frac{M_H^2}{M_{H^+}^2},x\right) - \mathcal{G}\left(\frac{M_{H^+}^2}{M_W^2},\frac{M_H^2}{M_W^2},x\right)\right].
\eea
\eesub
Finally, Fig.\ref{f:delmu_twoloop_inert} depicts the contributions stemming from the inert scalars via the $h\gamma\gamma$, $H\gamma\gamma$ and $H^+ W^- \gamma$ vertices. 
\besub
\bea
{\Delta a_\mu}_{\{\eta^+,~\phi \gamma\gamma \}}^{(2\text{loop})} &=&  \sum_{\phi = h, H} \frac{\alpha_{\text{em}} M_\mu^2}{8 \pi^3 M_{\phi}^2}~ \xi_\mu^{\phi}~ \lambda_{\phi \eta^+ \eta^-}\mathcal{F}^{(2)}\left(\frac{M_{\eta^+}^2}{M_{\phi}^2}\right), \\
{\Delta a_\mu}_{\{\eta_R,~H^+ W^-\gamma \}}^{(2\text{loop})} &=& \frac{\alpha_{\text{em}} M_\mu^2}
{64 \pi^3 s_w^2 (M_{H^+}^2 - M_W^2)}  
\xi_\mu^A  ~ \lambda_{ H^+ \eta^- \eta_R} \int_{0}^{1} dx~x^2 (x-1) \nonumber \\
&&\times \left[\mathcal{G}\left
(\frac{M_{\eta^+}^2}{M_{H^+}^2},\frac{M_{\eta_R}^2}{M_{H^+}^2},x\right) - \mathcal{G}\left(\frac{M_{\eta^+}^2}{M_W^2},\frac{M_{\eta_R}^2}{M_W^2},x\right)\right] \\ 
{\Delta a_\mu}^{2\text{loop}}_{\{\eta_I,~H^+ W^-\gamma \}}&=& \frac{\alpha_{\text{em}} M_\mu^2}{64 \pi^3 s_w^2 (M_{H^+}^2 - M_W^2)} \xi^A_\mu  ~ 
\lambda_{ H^+ \eta^- \eta_I} \int_{0}^{1} dx~x^2 (x-1) \nonumber \\
&&\times \left[\mathcal{G}\left(\frac{M_{\eta^+}^2}{M_{H^+}^2},\frac{M_{\eta_I}^2}{M_{H^+}^2},x\right) - \mathcal{G}\left(\frac{M_{\eta^+}^2}{M_W^2},\frac{M_{\eta_I}^2}{M_W^2},x\right)\right].
\eea
\eesub
The functions $\mathcal{F}^{(1)}(x),\mathcal{\tilde{F}}^{(1)}(x),\mathcal{F}^{(2)}(x)$ and $\mathcal{G}(a,b,x)$ are expressed in the Appendix.
To test the relative magnitudes of the different amplitudes, we compare their numerical values 
for the reference parameter point tan$\beta$ = 50, $M_H=M_{H^+}$ = 150 GeV while varying $M_A$. We first plot the one-loop contributions and the two-loop BZ contribution ${\Delta a_\mu}_{\{f,~\phi \gamma\gamma\}}^{(2\text{loop})}$ in Fig.\ref{f:contri1}.
One notes that the $A$-mediated one-loop amplitude is negative with a sizeable magnitude. In contrast, the two-loop amplitude
${\Delta a_\mu}_{\{\tau,A \gamma\gamma\}}$ is positive. And despite the additional loop suppression, it in fact dominates over the one-loop contribution owing to the 
$\frac{M^2_\tau}{M_\mu^2}$ enhancement factor. One must remember that though the amplitude involving $A$ and the $t$-quark implies multiplication by $\frac{M^2_t}{M_\mu^2}$, the same is proportional to cot$^2\beta$ due to the Yukawa scale factors.
Therefore, $(A,\tau)$ two-loop amplitude clearly beats the one from $(A,t)$ by a factor tan$^4\beta~\frac{M^2_\tau}{M_t^2} \simeq 650$. An $\sim \mathcal{O}(10^{-9})$ contribution to 
$\Delta a_\mu$ can therefore be induced for high tan$\beta$ and low $M_A$ thereby paving way for a resolution of the anomaly. We mention here that the contributions ${\Delta a_\mu}_{\{f,~H^+ W^- \gamma\}}^{(2\text{loop})}$, ${\Delta a_\mu}_{\{H^+,~\phi\gamma\gamma\}}^{(2\text{loop})}$ and ${\Delta a_\mu}_{\{H,H^+W^-\gamma\}}^{(2\text{loop})}$ are suppressed compared to ${\Delta a_\mu}_{\{\tau,A \gamma\gamma\}}$
in the parameter region of interest. More details about the purely 2HDM contribution can be found in~\cite{Broggio:2014mna,Cao:2009as,Wang:2014sda,Ilisie:2015tra,Abe:2015oca,Chun:2016hzs,Cherchiglia:2016eui,Dey:2021pyn}.

\begin{figure}
\centering
\subfigure[]{
\includegraphics[scale=0.47]{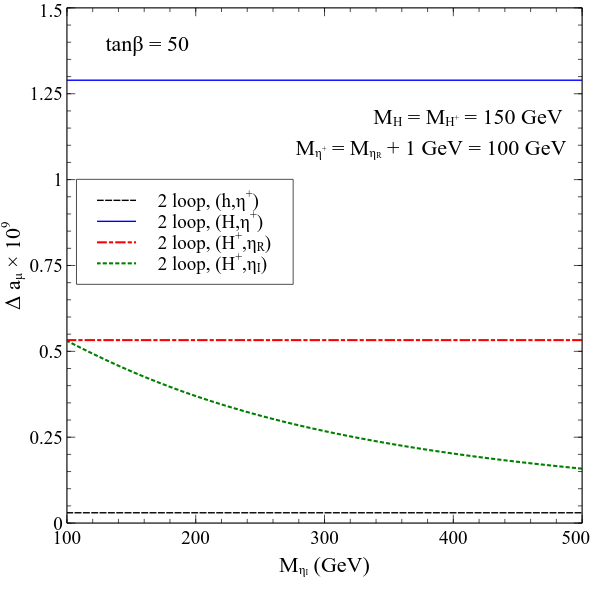}}
\subfigure[]{
\includegraphics[scale=0.47]{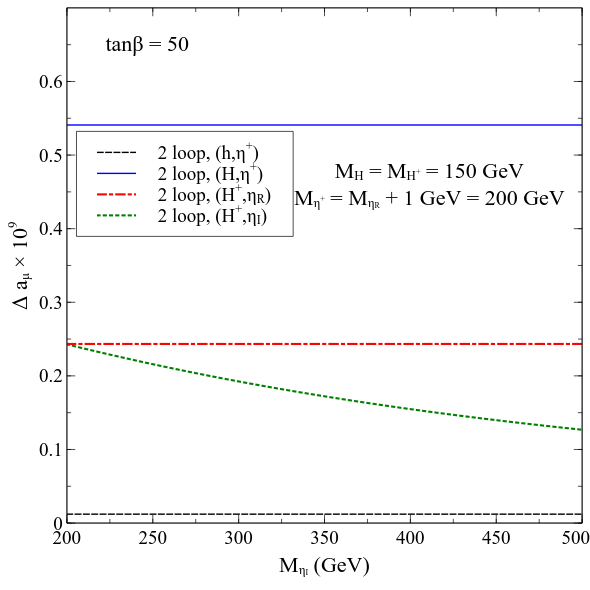}}
\caption{Two-loop Barr-Zee amplitudes induced by the inert scalars. The left (right) panel corresponds to $M_{\eta^+}$ = 100 GeV (200 GeV). The legends indicate the particles running in the loops. All trilinear couplings involving the inert scalars are chosen to be 2$\pi$ for illustration.}
\label{f:contri3}
\end{figure}

The more important component of the present discussion is obviously the contribution coming from the inert sector.
The trilinear couplings entering these amplitudes are expressed in the Appendix.
 Let us examine the trilinear coupling $\l_{H \eta^+ \eta^-}$ more closely. For $\alpha = \beta - \frac{\pi}{2}$, the dominant behaviour is 
$\l_{H \eta^+ \eta^-} \simeq \sigma_1 (c^2_\b - s^2_\b) \simeq \sigma_1$
for tan$\beta \gtrsim 20$. Therefore, $\l_{H \eta^+ \eta^-} \sim \mathcal{O}(1)$ can lead to large values of ${\Delta a_\mu}_{\{\eta,~H \gamma\gamma \}}^{(2\text{loop})}$. A similar argument reveals that $\l_{H^+ \eta^- \eta_R}$ and $\l_{H^+ \eta^- \eta_I}$ can also take sizeable values while $\l_{h \eta^+ \eta^-}$ remains suppressed for large tan$\beta$. Of course, the enhanced Yukawa scale factors of the $\mu$-lepton with $H,A,H^+$ also play a role here. We illustrate the strengths of the different $g-2$ amplitudes stemming from the inert scalars in Fig.\ref{f:contri3} for $M_{\eta_R}$ = 99 GeV and 199 GeV. We further fix $M_{\eta^+} = M_{\eta_R} + 1$ GeV and let $M_{\eta_I}$ vary. The magnitudes of $\l_{H \eta^+ \eta^-}$, $\l_{H^+ \eta^- \eta_R}$ and $\l_{H^+ \eta^- \eta_I}$ are each chosen to be $2\pi$ while $\l_{h \eta^+ \eta^-}$ = 1 is taken. The rationale behind such a choice is that trilinear couplings involving $H,A,H^+$ in one external leg and inert scalars in the other two can attain the maximum strength of $2\pi$ while remaining allowed by perturbative unitarity. Taking a common value for these couplings enables
a straightforward comparison of the relative magnitudes of the corresponding $g-2$ amplitudes. 
The largest contributor is the diagram involving the $H\gamma\gamma$ effective vertex. In fact, this amplitude alone adds $\simeq 1.25 \times 10^{-9}$ to the muon magnetic moment when $M_{\eta^+}$ = 100 GeV, as seen in Fig.\ref{f:contri3}(a).
An $\mathcal{O}(10^{-10})$ value is also generated for the same $M_{\eta^+}$ by the diagrams involving the $H^+ W^- \gamma$ effective interaction for the shown range of $M_{\eta_I}$. The inert sector thus alone suffices to resolve muon $g-2$ at 2$\sigma$ for $M_{\eta^+}$ = 100 GeV. Fig.\ref{f:contri3}(b) shows that the corresponding $g-2$ amplitudes are though expectedly smaller for $M_{\eta^+}$ = 200 GeV. In all, the two-loop amplitude driven by the inert scalars is substantial and more importantly, independent of $M_A$.

\begin{figure}
\centering
\subfigure[]{
\includegraphics[scale=0.48]{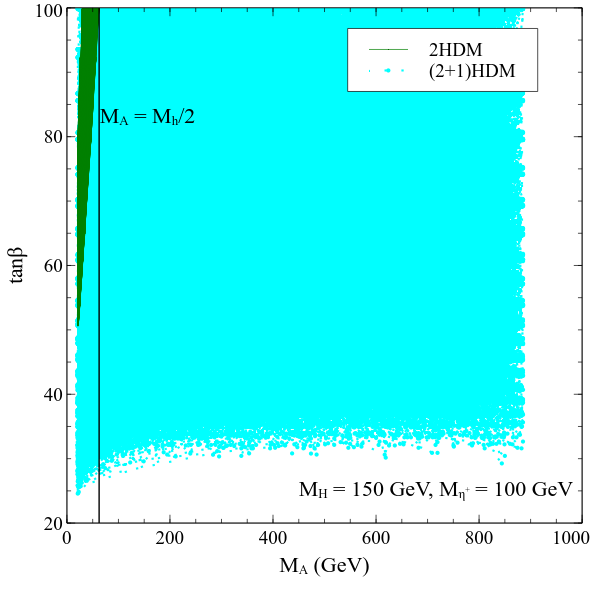}}
\subfigure[]{
\includegraphics[scale=0.48]{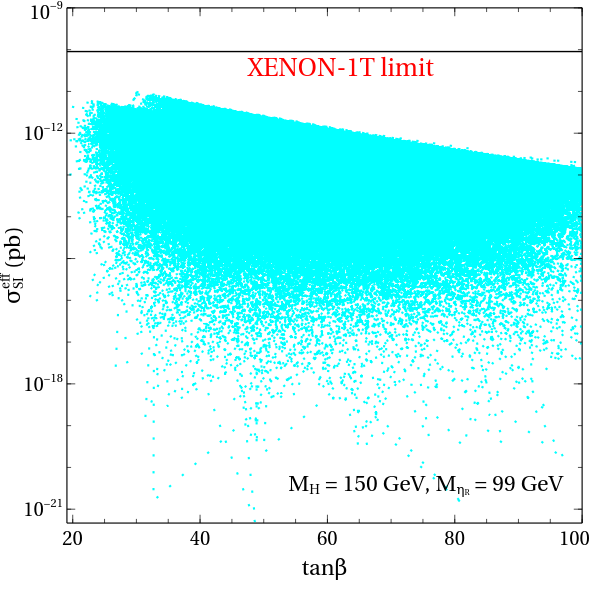}} 
\caption{(a) Parameter space in the $M_A-$tan$\beta$ plane compatible with 
the observed $\Delta a_\mu$ at 2$\sigma$ for $M_{\eta^+}=M_{\eta_R}$ + 1 GeV = 100 GeV. The region to the left of the vertical line is tightly constrained by BR($h \to A A$) measurements.
(b) Prediction of $\sigma^{\text{eff}}_{\text{SI}}$ versus $\tan\beta$ for the parameter points in the (2+1)HDM compatible with $\Delta a_\mu$ at 2$\sigma$ for $M_{\eta^+}=M_{\eta_R}$ + 1 GeV = 100 GeV. The color-coding is explained in the legends. }
\label{f:MA-tb_ls_100}
\end{figure}

\begin{figure}
\centering
\subfigure[]{
\includegraphics[scale=0.48]{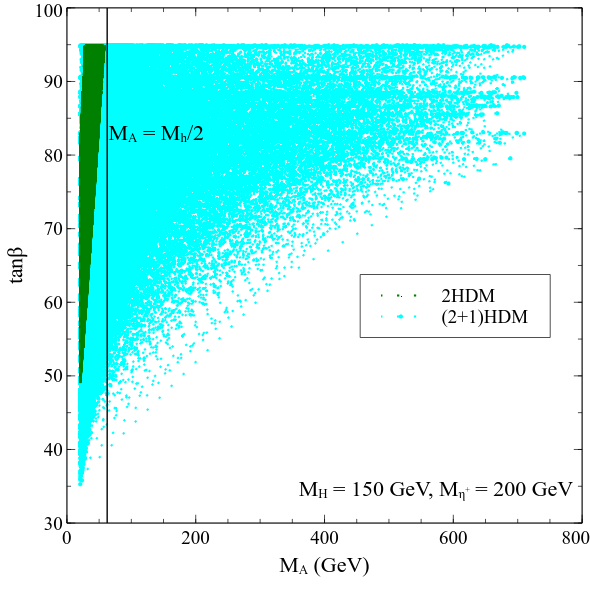}}
\subfigure[]{
\includegraphics[scale=0.48]{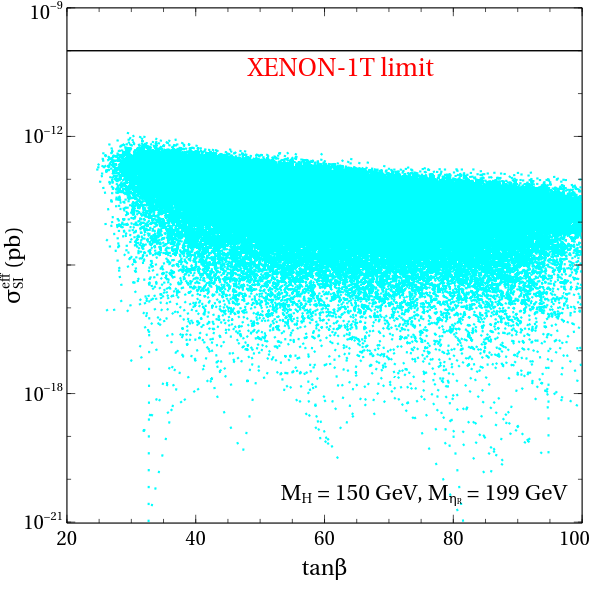}} 
\caption{(a) Parameter space in the $M_A-$tan$\beta$ plane compatible with 
the observed $\Delta a_\mu$ at 2$\sigma$ for $M_{\eta^+}=M_{\eta_R}$ + 1 GeV = 200 GeV. The region to the left of the vertical line is tightly constrained by BR($h \to A A$) measurements.
(b) Prediction of $\sigma^{\text{eff}}_{\text{SI}}$ versus $\tan\beta$ for the parameter points in the (2+1)HDM compatible with $\Delta a_\mu$ at 2$\sigma$ for $M_{\eta^+}=M_{\eta_R}$ + 1 GeV = 200 GeV. The color-coding is explained in the legends. }
\label{f:MA-tb_ls_200}
\end{figure}

We propose a numerical scan to this end to validate the model against the constraints. And this brings us to a counting of the independent model parameters. In the 2HDM sector, $m_{11}$ and $m_{22}$ are eliminated by the tadpole conditions: $\frac{\partial{V}}{\partial{v_1}}$ = $\frac{\partial{V}}{\partial{v_2}}$ = 0. The couplings $\l_{1-5}$ can be expressed in terms of the physical masses, mixing angles and 
$\l_{6,7}$. In the inert sector, we define the parameters $\l_{L_1}$ and $\l_{L_2}$ below in lines similar to the IDM.
\besub
\bea
\l_{L_1} &=& \nu_1 + \omega_1 + k_1, \\
\l_{L_2} &=& \nu_2 + \omega_2 + k_2.
\eea
\eesub
For insight, the $\eta_R-\eta_R-h$ and $\eta_R-\eta_R-H$ couplings are linear combinations of $\l_{L_1} v$ and $\l_{L_2} v$ when $\sigma_1,\sigma_2,\sigma_3=0$. The parameters $\mu,\omega_2,k_2,\nu_1,\nu_2$
can be traded off using the relations
\besub
\bea
\mu^2 &=& M^2_{\eta_R}
- \frac{1}{2} \l_{L_1} v^2 c^2_\b
- \frac{1}{2} \l_{L_2} v^2 s^2_\b
- (\sigma_1 + \sigma_2 + \sigma_3)v^2 
s_\b c_\b,\\
\omega_2 &=& \frac{M^2_{\eta_R} + M^2_{\eta_I} - 2 M^2_{\eta^+} + \omega_1 v^2 c^2_\b
- 2 \sigma_2 v^2 s_\b c_\b}{v^2 s^2_\b}, \\
k_2 &=& \frac{M^2_{\eta_R} - M^2_{\eta_I} 
- k_1 v^2 c^2_\b - 2 \sigma_3 v^2 s_\b c_\b}{v^2 
s^2_\b}, \\
\nu_1 &=& \l_{L_1} - \omega_1 - k_1, \\
\nu_2 &=& \l_{L_2} - \omega_2 - k_2.
\eea
\eesub
With the alignment limit in place, the independent parameters in the (2+1)HDM are therefore 
$\{m_{12}, M_H, M_A, M_{H^+}, M_{\eta_R}, M_{\eta_I}, M_{\eta^+}, 
\text{tan}\beta, \l_6, \l_7, \omega_1, \k_1, \s_1, \s_2, \s_3, \l_{L_1}, \l_{L_2} \}$. We adhere to the same choices for $M_H,M_{H^+},M_{\eta^+},M_{\eta_R}$ as in the preceding discussion. Additionally, we take $\l_6 = \l_7 = \l_{L_1} = \l_{L_2} = 0.01$. The choice of the 2HDM masses is consistent with various
exclusion constraints~\cite{Chowdhury:2017aav} from the LHC on account of the suppressed couplings to quarks in the Type-X case.  The rest of the parameters are varied as follows.
\besub
\bea
0 < m_{12} < 1~\text{TeV},~~20~\text{GeV} < M_A < 1~\text{TeV}, \nonumber \\
M_{\eta_R} + 1~\text{GeV} \leq M_{\eta_I} \leq 500~\text{GeV}, \nonumber \\ 
10 < \text{tan}\beta < 100, 
~~~|\omega_1|,~|\k_1| < 4\pi, \nonumber 
~~|\s_1|,~|\s_2|,~|\s_3| < 2\pi.
\eea
\eesub 
The (minimum) 1 GeV mass-splitting $\eta_R$ has with $\eta_I$ and $\eta^+$   disallows $W,Z$-mediated inelastic direct detection scatterings~\cite{arina:2009}. Such mass-gaps are also consistent with the $\Delta T$ constraint. The parameter points passing all the constraints are plotted in the 
$M_A-$ tan$\beta$ plane in 
Figs.\ref{f:MA-tb_ls_100}(a) and \ref{f:MA-tb_ls_200}(a). The most important finding to emerge is that the parameter space compatible with the observed muon $ g-2$ excess appreciably expands in presence of an additional inert scalar doublet. Fig.\ref{f:MA-tb_ls_100}(a) shows that an $A$ as heavy as 800 GeV is now allowed for tan$\beta \simeq$ 35 for $M_{\eta^+}$ = 100 GeV. This enhancement is clearly attributed to the BZ contributions induced by the inert scalars. Though the enhancement is less in case of $M_{\eta^+}$ = 200 GeV, an $M_A$ = 250 GeV still complies with $\Delta a_\mu$ for tan$\beta \simeq$ 55 (see Fig.\ref{f:MA-tb_ls_200}(b)).

Figs.\ref{f:MA-tb_ls_100}(b) and \ref{f:MA-tb_ls_200}(b) display $\sigma_{\text{SI}}^{\text{eff}}$ versus tan$\beta$ for the same parameter points. The (2+1)HDM features (co)annihilations mediated by the 2HDM scalars. For instance, $\eta_R \eta_R \to f \bar{f}$ mediated by an s-channel $H$ is an important annihilation channel, where $f$ denotes a SM fermion. This is not encountered in the case where only one active Higgs doublet is present.
Since the $M_{\eta^+} - M_{\eta_R}$ = 1 GeV in this work, $\eta^+ \eta_R \to f \bar{f^\prime}$ and $\eta^+ \eta^- \to f \bar{f}$ coannihilations are also triggered mediated by s-channel $H^+$ and $H$ respectively. And the most dominant fermionic co(annihilations) are to $\tau\bar{\tau}$ and $\tau\bar{\nu_\tau}$ on account of the tan$\beta$-enhanced Yukawa interactions. The corresponding co(annihilation) cross sections are thus quite large thereby leading to a small $\Omega h^2$. Consequently, $\sigma^{\text{eff}}_{\text{SI}}$ remains below the stipulated bound for all the parameter points. This is confirmed by 
Figs.\ref{f:MA-tb_ls_100}(b) and \ref{f:MA-tb_ls_200}(b). A more detailed discussion of DM phenomenology in the (2+1)HDM setup is beyond the scope of this study and can be taken up as a future endeavour.

\section{LHC signatures}\label{collider}

We discuss in this section a prospective collider signature of the (2+1)HDM framework at the 14 TeV LHC. For $M_A < \frac{M_h}{2}$, a useful channel to look for the $A$ in the pure Type-X is $p p \to h \to A A \to \tau^+ \tau^- \tau^+ \tau^-,~\tau^+ \tau^- \mu^+ \mu^-$~\cite{Chun:2017yob,Chun:2018vsn}. The large $h$-production cross section through gluon-fusion leads to a healthy event rate. However, The primary focus of this work is on a heavier $A$ for which $h \to A A$ is kinematically closed. Hence, the aforementioned channel is not suited to our case. The channel $p p \to H \to A A$, though kinematically open in principle, is also not promising for a couple of reasons. First, this channel mandates $M_H > 2 M_A$. Such a heavy $H$ would diminish the relevant BZ amplitudes accordingly. Secondly, the $p p \to H$ production cross section is small owing to the suppressed Yukawa couplings involved here. The selection of LHC signals
for the (2+1)HDM is guided by the two expectations: (a) the most promising fermionic decay channel is $A \to \tau^+ \tau^-$ on account of the enhanced Yukawa coupling. Further, the $\tau$-pair will be more boosted in case of an heavier $A$. (b) Involving the inert scalars in the signals should ultimately lead to a modified $\met$ spectrum \emph{w.r.t.} the SM which in turn could be a distinguishing kinematical feature. In view of this, we propose to look at the $p p \to \eta_R \eta_A \to \eta_R \eta_R A \to \tau^+ \tau^- + \met$ signal. We give the details of the analysis below.

The $\eta_R \eta_I$ pair is dominantly produced via an s-channel exchange of $Z$. A small contribution also comes from $A$-exchange. Following an $\eta_I \to \eta_R A$ decay, 
the $A$ subsequently decays to a $\tau^+\tau^-$ pair. 
The lightest inert scalar $\eta_R$ registers as missing transverse momentum ($\met$). Thus, a modified $\met$ spectrum \emph{w.r.t.} the SM can be a potential handle to discern the signal from the background. We also look for completely hadronic decays of the $\tau$-pair thereby leading to a $2 \tau_h + \met$ signature. Instead of a conventional cut-based analysis, we intend to analyse this signal using the more advanced multivariate techniques. A few benchmark points (BPs) are put forth in in Table \ref{tab:BP}. The BPs satisfy all the applied constraints and predict the requisite value of $\Delta a_\mu$, as can be read from Table \ref{tab:BP}. All the three BPs are characterised by $M_{\eta_I} > M_{\eta_R} + M_A$. For these BPs, appropriate values of the quartic couplings ensure that $\eta_I \to \eta_R A$ is the leading decay mode. The subleading one is in fact $\eta_I \to \eta_R Z$. When it comes to the decay of $A$, it is observed that $A \to Z H, W^\pm H^\mp$ can compete with $A \to \tau^+ \tau^-$ for the choice of $M_H = M_{H^+}$ = 150 GeV. This is found true especially for BP3. However, the $A \to \tau^+ \tau^-$ branching fraction in still $\mathcal{O}(10\%)$ nonetheless.

\begin{table}
\centering
\begin{tabular}{ |c|c|c|c| } 
\hline
& BP1 & BP2 & BP3 \\ \hline
$m_{12}$ & 24.0 GeV & 20.4 GeV & 21.6 GeV \\
tan$\b$ & 38.71 & 53.83 & 47.98 \\
$M_A$ & 206.2 GeV & 253.24 GeV & 301.26 GeV \\
$M_{\eta_I}$ & 346 GeV & 397 GeV & 450.5 GeV \\
$k_1$ & -0.992743 & -2.07345 & -0.55292 \\
$\omega_1$ & -2.94053 & -0.125664 & 0.13823 \\
$\s_1$ & -5.00142 & -5.70513 & -6.09469 \\
$\s_2$ & 5.7554 & -0.263894 & 1.29434 \\
$\s_3$ & 4.05894 & 5.44124 & 5.90619 \\
$\Delta a_\mu$ & 1.48646 & 1.51138 & 1.64289 \\
$\sigma^{eff}_{SI}$ & $5.28 \times 10^{-48}$ cm$^2$ & $3.81 \times 10^{-50}$ cm$^2$ & $4.42 \times 10^{-49}$ cm$^2$ \\ \hline
BR$(\eta_I \to \eta_R A)$ & 0.844604 & 0.822958 & 0.748021 \\ 
BR$(A \to \tau^+ \tau^-)$ & 0.99 & 0.7983 & 0.341914 \\ \hline
\end{tabular}
\caption{BPs used to study the discovery prospects of an $A$ in the (2+1)HDM (lepton-specific). The values for the rest of the masses are 
$M_H = M_{H^+} = 150$ GeV, $M_{\eta^+} = M_{\eta_R} + 1$ GeV = 100 GeV.}
\label{tab:BP}
\end{table}

\begin{table}
\centering
\begin{tabular}{ |c c c| } 
\hline
Signal/Backgrounds & Process & Cross section \\ \hline
Signal & & \\
BP1 & & 7.904 fb \\
BP2 & $p p \to \eta_R \eta_I \to \eta_R \eta_R A \to \tau^+\tau^- + \met$
 &  3.867 fb \\ 
BP3 & & 0.965 fb \\ \hline
Backgrounds & & \\
 & $pp \to j j + \met$ & $1.11 \times 10^6$ fb \\
 & $pp \to \tau^+ \tau^- + \met$ & $5.80 \times 10^2$ fb \\
 & $pp \to t\bar{t} \to b\bar{b}W^+ W^- \to \tau^+ \tau^- b \bar{b} + \met$ & $8.092 \times 10^3$ fb (NNLO) \\
  & $pp \to W^\pm Z \to \tau^+ \tau^- \tau^\pm + \met$ & $4.31 \times 10^1$ fb \\ \hline
\end{tabular}
\caption{Signal and background cross sections at the 14 TeV LHC.}
\label{tab:cs}
\end{table}

The relevant interactions of the model have been incorporated in \texttt{FeynRules}\cite{Alloul:2013bka}.
We next discuss the possible background processes. The largest background comes from $p p \to j j + \met$ ($j$ denotes a light jet) when both $j$s are misidentified as $\tau_h$s. As the $j j + \met$ cross section is $\sim 10^6$ fb, even a small misidentification rate leads to a large $2\tau_h + \met$ cross section. Another important background is $p p \to \tau^+ \tau^- + \met$ that mainly comes from $p p \to W^+ W^-,~Z Z$ production. Owing to the large $p p \to t \bar{t}$ cross section, the $p p \to t \bar{t} \to \tau^+ \tau^- b \bar{b} + \met$ process can also lead to a sizeable background when both the 
$b$-jets are missed. A small contribution also comes from $p p \to W^\pm Z \to \tau^+ \tau^- \tau^\pm + \met$ when one $\tau$ is missed. The cross sections of the signal BPs and the backgrounds are given in Table \ref{tab:cs}. We have used \texttt{MG5aMC@NLO}~\cite{Alwall:2014hca} to generate the signal and background events at the leading order (LO). The CTEQ6L Parton Distribution Function (PDF) set and default hadronization and factorization scales are used. The parton level events are passed on to \texttt{pythia8}~\cite{Sjostrand:2014zea} for showering and hadronization and subsequently to \texttt{Delphes-3.4.1}~\cite{deFavereau:2013fsa} for detector simulation. Specifically, we have throughout used the default CMS detector simulation card that comes with \texttt{Delphes-3.4.1}. For an integrated luminosity $L$, the number in a signal or background sample of events is determined as 
$L \times \sigma \times \epsilon$ with $\sigma$ and $\epsilon$ respectively referring to the cross section and cut-efficiency. The signal significance is computed using the formula
$S = \sqrt{2\Big[(N_S + N_B)\text{log}\big(\frac{N_S + N_B}{N_B}\big)
 - N_S\Big]}$~\cite{Cowan:2010js} where $N_S$ ($N_B$) denotes the number of signal (background) events. 
 
\begin{figure}
\centering
\subfigure[]{
\includegraphics[scale=0.48]{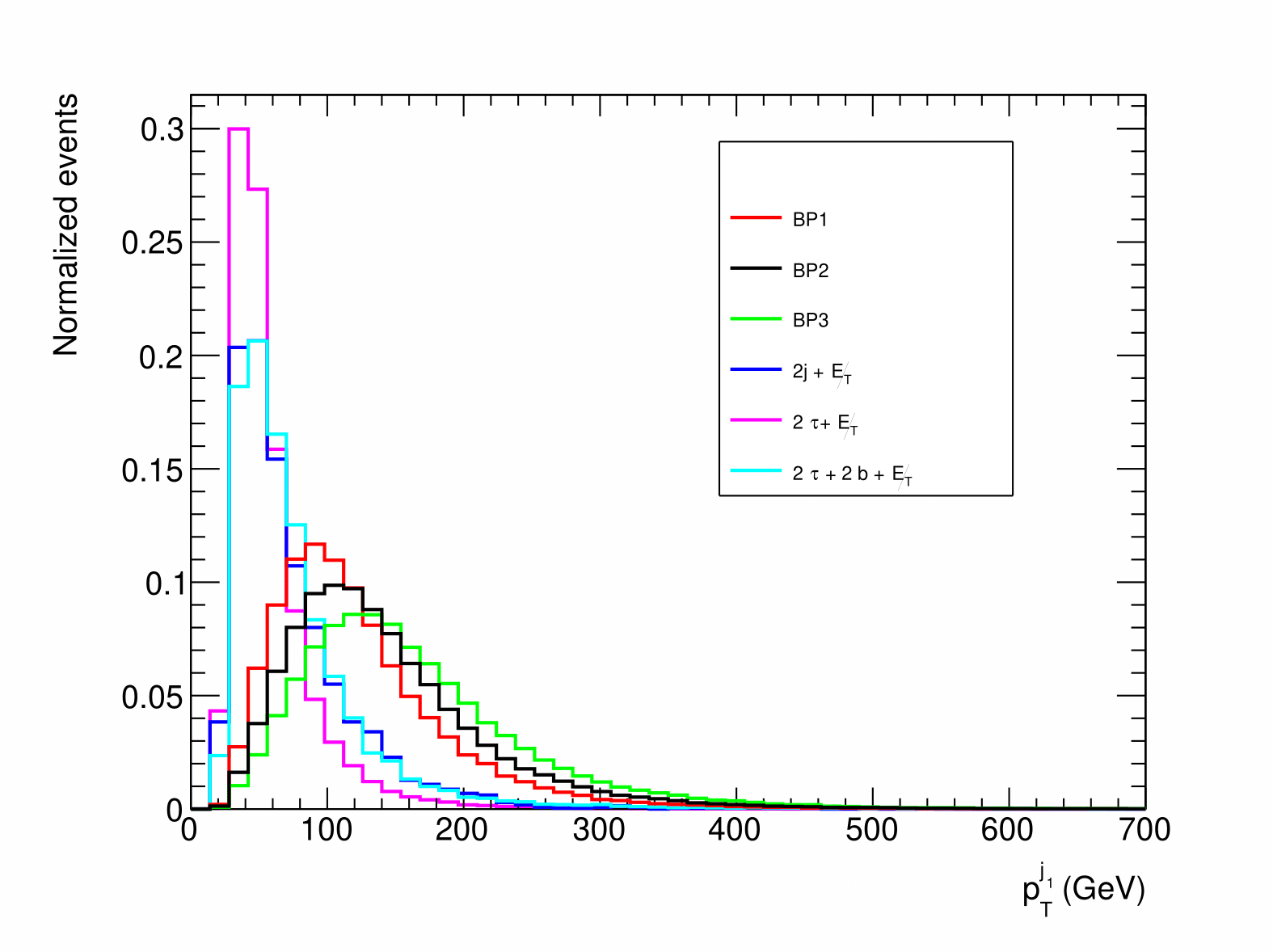}}
\subfigure[]{
\includegraphics[scale=0.48]{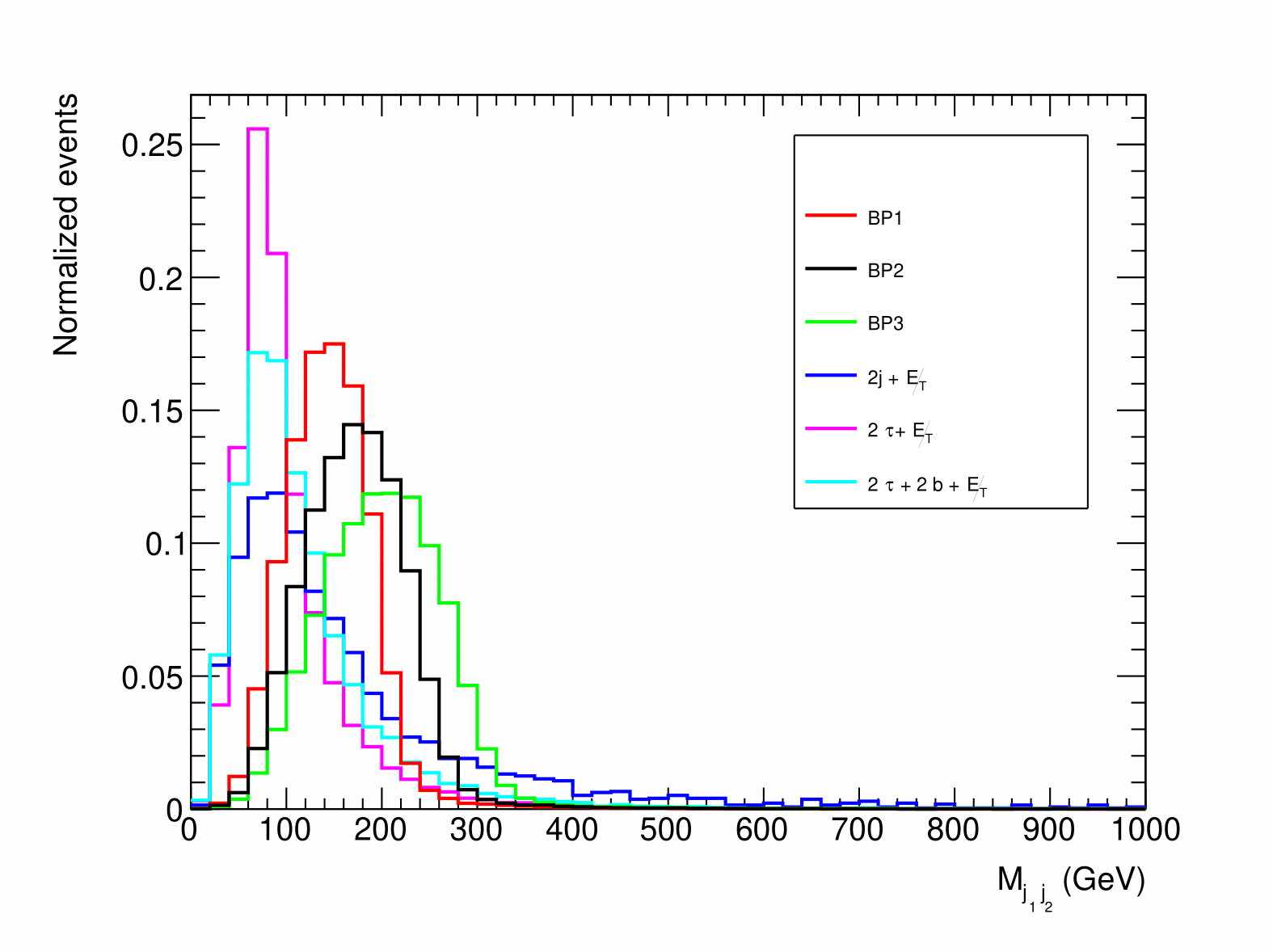}}
\caption{The distributions of $p_T^{j_1}$ and $M_{j_1j_2}$ in panels (a) and (b) respectively. The colour coding is given in the legends.}
\label{f:distn1}
\end{figure}

\begin{figure}
\centering
\subfigure[]{
\includegraphics[scale=0.48]{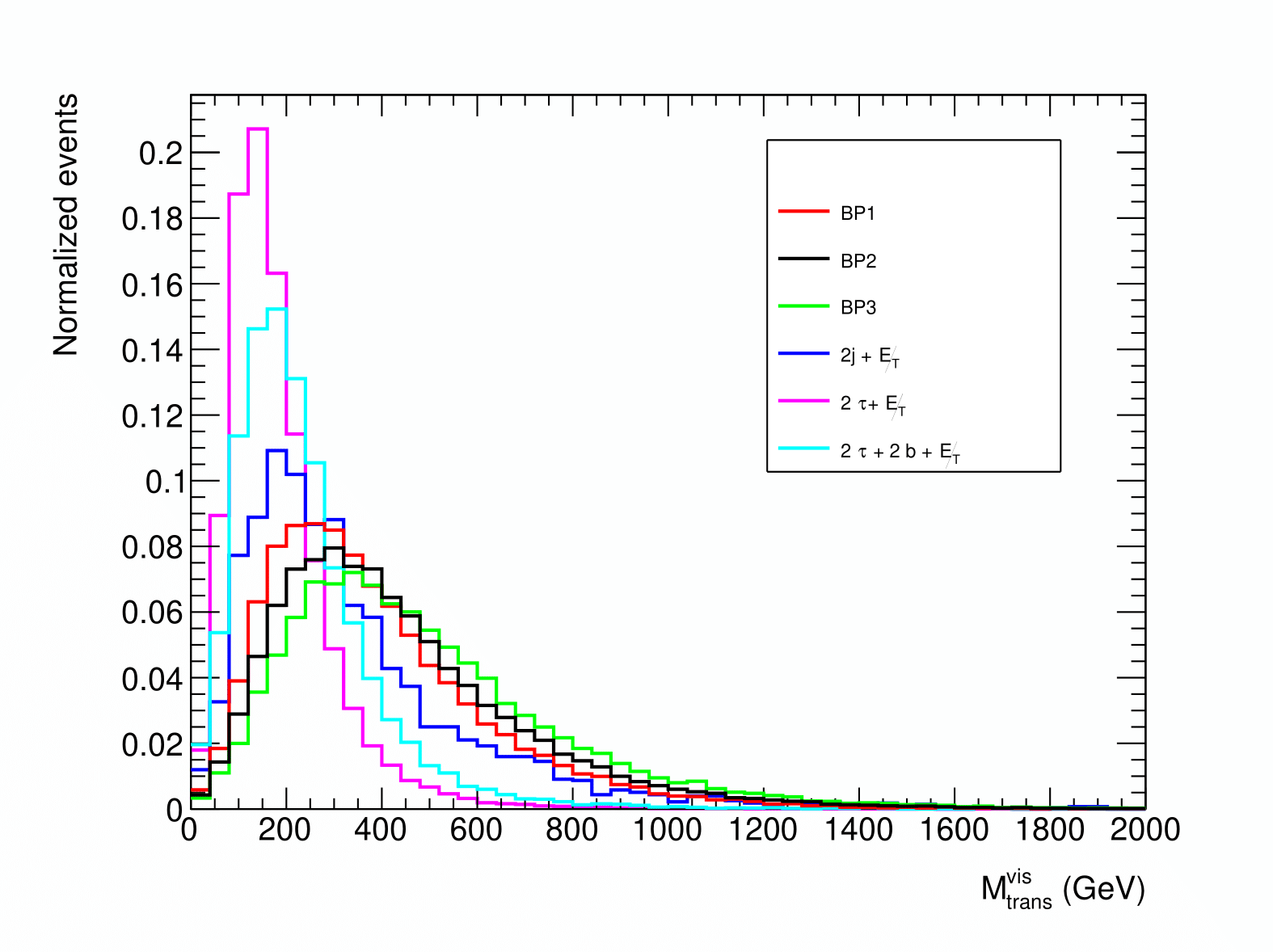}}
\subfigure[]{
\includegraphics[scale=0.48]{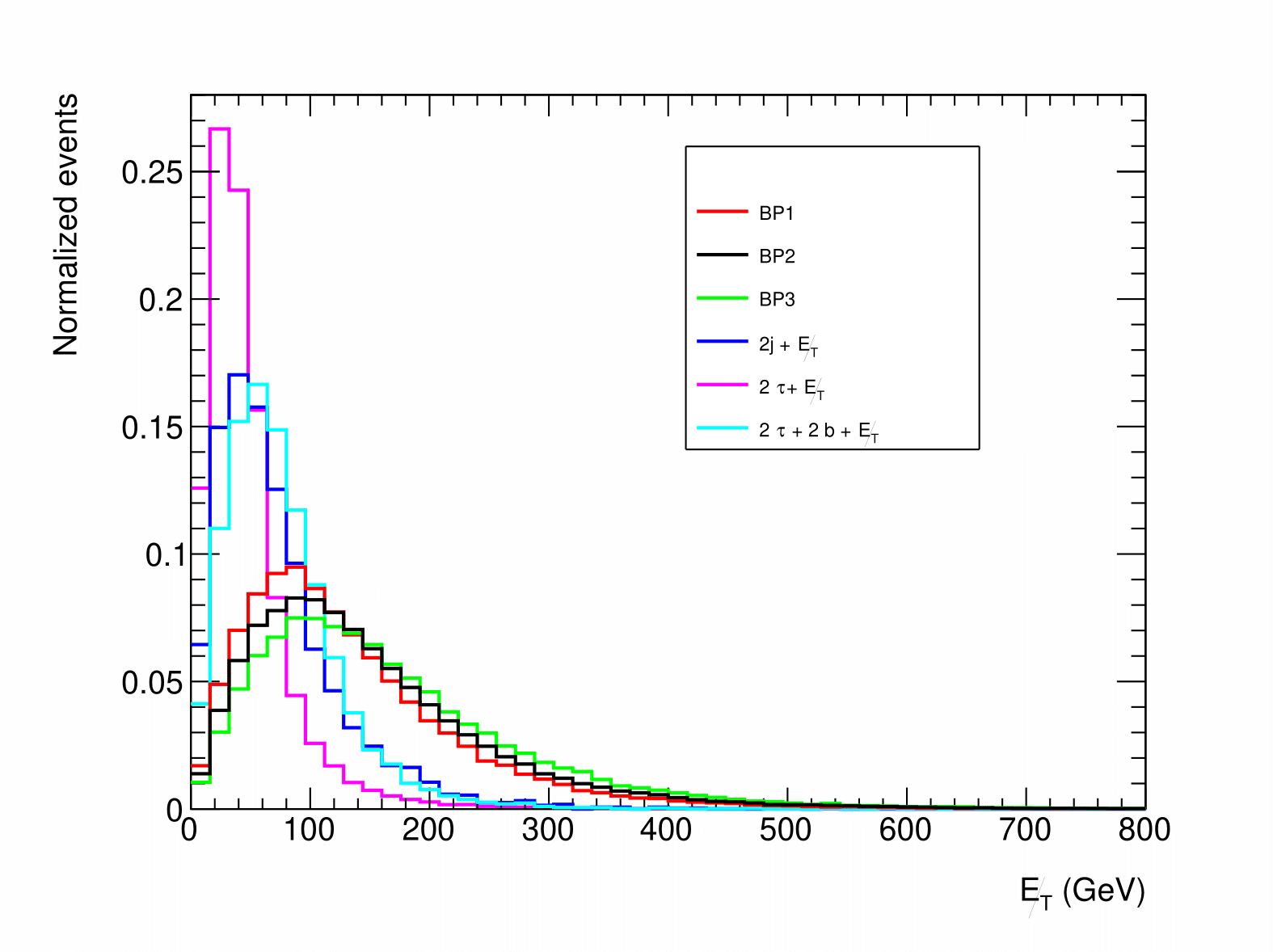}}
\caption{The distributions of $M^{\text{vis}}_{\text{trans}}$ and $\met$ in panels (a) and (b) respectively. The colour coding is given in the legends.}
\label{f:distn2}
\end{figure}

Events are selected by demanding exactly two $\tau_h$ and zero $b$-jets. Sizeable background fractions are eliminated at this level itself since the background processes in this case mostly lead to flavor-democratic leptons.
In addition to this demand, the following trigger-level cuts are also applied:
\bea
p_T^\ell > 10~\text{GeV},~|\eta_{j,\ell}| < 2.5,~\Delta R_{\ell\ell} > 0.2,~\Delta R_{\ell j} > 0.2,~\Delta R_{jj} > 0.4.
\eea
Inthe above, $\Delta R_{mn} = \sqrt{\Delta \eta^2_{mn} + \Delta \phi^2_{mn}}$ and $\Delta \eta_{mn},\Delta \phi_{mn}$ are the differences between pseudorapidity and azimuthal angles of $m$th and $n$th particles respectively. In addition, $\ell = e,\mu$. It is reminded that light jets come only from showering for the signal and  backgrounds for this analysis. We denote the two leading $\tau$-tagged jets by $j_1$ and $j_2$ in the decreasing order of their $p_T$. The following kinematic variables of interest are identified; $p_T^{j_1}$: the transverse momentum of the leading tau-jet, $\met$: the missing transverse energy, $M_{j_1 j_2}$: nnvariant mass of the $\tau_h\tau_h$ pair, and, $M^{\text{vis}}_{\text{trans}}$: transverse mass of the $\tau_h\tau_h$ pair. The normalised distributions of these four variables are shown in Figs.\ref{f:distn1} and \ref{f:distn2}. A brief discussion on the kinematic features is in order.  The distribution of $p_T^{j_1}$ is shown in Fig.\ref{f:distn1}(a) for the signal BPs and the backgrounds. The $p_T^{j_1}$ spectrum is seen to be \emph{harder} in case of the signals than the backgrounds. In fact, the heavier the pseudoscalar, the more boosted are the $\tau_h\tau_h$ pair and 
hence, the harder is the $p_T^{j_1}$ spectrum. One inspects that the distribution peaks around $\simeq$ 100 GeV, 120 GeV and 120 GeV for BP1, BP2 and BP3 respectively. On the other hand, the backgrounds have their peaks below 100 GeV. 
Next, Figs.\ref{f:distn1}(b) and \ref{f:distn1}(c) show that the spectra of the invariant mass and transverse mass of the $\tau_h\tau_h$ pairs share a correlation. That is, the larger the $M_A$, the higher is the value where these distributions peak. One must however note that the peak of the $M_{j_1 j_2}$ distribution cannot coincide with $M_A$ on account of the $\met$ component in $\tau$ decays.

We briefly discuss the sources of $\met$ in the signal and backgrounds. In case of the $p p \to W^+ W^-, Z Z$ backgrounds, for instance, neutrinos coming from $W^\pm \to \tau^\pm \nu$ and $Z \to \bar{\nu} \nu$ majorly decide the shape of the $\met$ distribution. The $p p \to t \bar{t}$ production leads to neutrinos via the $W^\pm$ decays. A subleading effect for these backgrounds also comes from the neutrinos emerging from $\tau$ decays. In case of the signals however, the main source of $\met$ is in fact the inert scalar $\eta_R$. The shape of the $\met$ distribution for the signals thus depends on the transverse momentum of $\eta_R$. And the masses of $\eta_I$ and $A$ are such for the BPs that the distribution is harder than the backgrounds. The background distributions all peak below 100 GeV as opposed to the signals that peak at $\gtrsim$ 100 GeV, as seen in Fig.\ref{f:distn1}(d).

We now turn to the multivariate analysis (MVA) using Decorrelated Boosted Decision Tree (BDTD) algorithm as implemented within the Toolkit for Multivariate Data Analysis (TMVA)~\cite{2007physics3039H} framework. A brief overview of the method is as follows. To classify an events as signal-like or background-like, decision trees are used as classifiers. One discriminating kinematic variable with an optimised cut value applied on it is associated with each node of the decision tree, to make the best possible distinction between the signal-like and background-like events. The handle to do this
within TMVA is to tune the BDTD variable \texttt{NCuts}. The training of the decision trees starts from a zeroth node and continues till a particular depth specified by the user is reached. This particular depth is termed as \texttt{MaxDepth}. Finally from the final nodes or the leaf nodes, an event can be specified as signal or background according to their purity. An event can be tagged as signal (background) when $p > 0.5$ ($p < 0.5$).

The decision trees are considered weak classifiers as they are prone to statistical fluctuations of the training sample. To circumvent this problem, one can combine a set of weak classifiers into a stronger one and create new decision trees by modifying the weight of the events. This procedure is referred to as Boosting. In this analysis, we choose Adaptive boost with the input variables transforming in a decorrelated manner, since this is very useful for weak classifiers. It is implemented as Decorrelated AdaBoost in TMVA.  To avoid over training of the signal and background samples, the result of the Kolmogorov-Smirnov test, i.e. Kolmogorov-Smirnov score (KS-score) is demanded to be always > 0.01 and stable.

\begin{figure}
\centering
\includegraphics[scale=0.50]{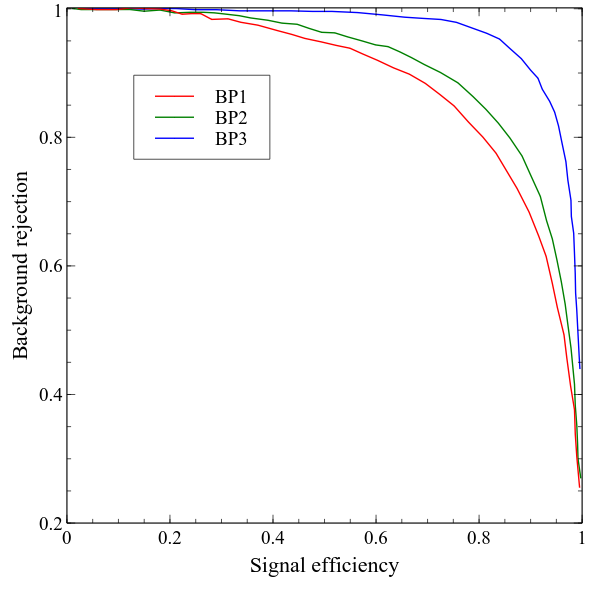}
\caption{The KS-scores for the chosen lepton-specific benchmarks. The top-left, top-right and bottom panels correspond to BP1, BP2 and BP3 respectively.}
\label{f:roc}
\end{figure}

Now the BDTD algorithm orders the kinematic variables that are fed into the algorithm by their importance in discriminating the signal from the background. The following kinematic variables are proposed in this analysis:
\bea
p_T^{j_1}, p_T^{j_2}, \eta_{j_1}, \eta_{j_2}, \phi_{j_1}, \phi_{j_2},
\Delta \phi_{j_1 \met},\Delta \phi_{j_2 \met},\Delta \phi_{j_1 j_2},\Delta \eta_{j_1 j_2},\Delta R_{j_1 j_2}, \met, M_{\text{eff}}, p^{j_1 j_2}_T, M_{j_1 j_2}, M_{\text{trans}}^{\text{vis}}.
\nonumber
\eea
In the above, $p_T^{j_1j_2}$ straightforwardly refers to the \emph{vector} transverse momentum of the $\tau_h\tau_h$ system, i.e., $p_T^{j_1j_2} \equiv \sqrt{(p_x^{j_1} + p_x^{j_2})^2
 + (p_y^{j_1} + p_y^{j_2})^2}$. The variables that turn out to be the most important in BDT ranking are $p_T^{\mu_1}, p_T^{\mu_2}, \eta_{\mu_1}, \eta_{\mu_2}, p_T^{\mu_1\mu_2}, \Delta R_{\mu_1\mu_2}, M^{\mu\mu}_{\text{inv}}, \met$. The relevant BDT parameters are tabulated in Table \ref{tab:BDT_param}. The signal and background
distributions along with their KS-scores are depicted in Fig.\ref{f:KS} in the Appendix.
%\begin{figure}
%\centering
%\includegraphics[scale=0.55]{ROC.png}
%\caption{The ROC curves for the chosen benchmarks. The color coding is given in the legends.}
%\label{f:ROC}
%\end{figure}
The degree of background rejection for each BP can be gauged from the Receiver's Operative Characteristic (ROC) curves as shown in Fig.\ref{f:roc}. The efficiency of background rejection is seen to improve sequentially from BP1 to BP3. And this trend is only expected on account of the progressively smaller overlap recorded between the signal BPs and the background upon going from BP1 to BP3, as also concurred by Fig.\ref{f:KS}. The yields at an integrated luminosity 3000 fb$^{-1}$ for the signal benchmarks and the backgrounds after optimisation through BDTD-analysis are given in Table \ref{tab:BDT_yield}.

\begin{table}
\centering
\begin{tabular}{ |c c c c c c c| } 
\hline
BP & \texttt{Ntrees} & \texttt{MinNodeSize} & \texttt{MaxDepth} & \texttt{NCuts} & KS-score for signal (background) & BDT Score \\ \hline
BP1 & 120 & 2.5 & 2 & 55 & 0.487 (0.035) & 0.3521 \\
BP2 & 150 & 3 & 2 & 40 & 0.439 (0.016) & 0.4047 \\
BP3 & 120 & 3 & 2 & 50 & 0.038 (0.105) & 0.5336\\ \hline
\end{tabular}
\caption{Tuned BDT parameters for BP1, BP2, BP3}
\label{tab:BDT_param}
\end{table}

\begin{center}
\begin{table}[htb!]
\centering
\scalebox{1.1}{%
\begin{tabular}{|c|c|c|}\hline
\multicolumn{3}{|c|}{BP1} \\ \hline
 & Process  & Yield at 3000 fb $^{-1}$ \\ \hline \hline
\multirow{4}{*}{Background}  
 & $j j + \met$             & $671$ \\  
 & $\tau^+ \tau^- + \met$             & $557$ \\ 
 & $\tau^+ \tau^-  b\bar{b} + \met$            & $3276$ \\ 
 & $\tau^+ \tau^- \tau^\pm + \met$            & $45$ \\ \cline{2-3}  
 & $N_B^{\text{BDT}}$                      & $4548$ \\ \hline
\multicolumn{2}{|c|}{$N_S^{\text{BDT}}$} & $481$ \\\hline 
\end{tabular}}
\quad
\scalebox{1.1}{%
\begin{tabular}{|c|c|c|}\hline
\multicolumn{3}{|c|}{BP2} \\ \hline
 & Process  & Yield at 3000 fb $^{-1}$ \\ \hline \hline
\multirow{4}{*}{Background}  
 & $j j + \met$             & $\sim 0$ \\   
 & $\tau^+ \tau^- + \met$             & $128$ \\ 
 & $\tau^+ \tau^-  b\bar{b} + \met$            & $1266$ \\ 
 & $\tau^+ \tau^- \tau^\pm + \met$            & $ 12$ \\ \cline{2-3}  
 & $N_B^{\text{BDT}}$                      & $1405$ \\ \hline
\multicolumn{2}{|c|}{$N_S^{\text{BDT}}$} & $177$ \\\hline 
\end{tabular}} \\ 
\quad \\
\scalebox{1.1}{%
\begin{tabular}{|c|c|c|}\hline
\multicolumn{3}{|c|}{BP3} \\ \hline
 & Process  & Yield at 3000 fb $^{-1}$ \\ \hline \hline
\multirow{4}{*}{Background}   
 & $j j + \met$             & $\sim 0$ \\  
 & $\tau^+ \tau^- + \met$             & $21$ \\ 
 & $\tau^+ \tau^-  b\bar{b} + \met$            & $273$ \\ 
 & $\tau^+ \tau^- \tau^\pm + \met$            & $3$ \\ \cline{2-3}  
 & $N_B^{\text{BDT}}$                      & $297$ \\ \hline
\multicolumn{2}{|c|}{$N_S^{\text{BDT}}$} & $19$ \\\hline 
\end{tabular}}
\bigskip
\caption{The signal and background yields at $3000~{\rm fb}^{-1}$ for BP1, BP2 and BP3  for the $\tau_h\tau_h + \met$ channel as obtained from the BDTD analysis.}
\label{tab:BDT_yield}
\end{table}
\end{center}

\begin{figure}
\centering
\subfigure[]{
\includegraphics[scale=0.48]{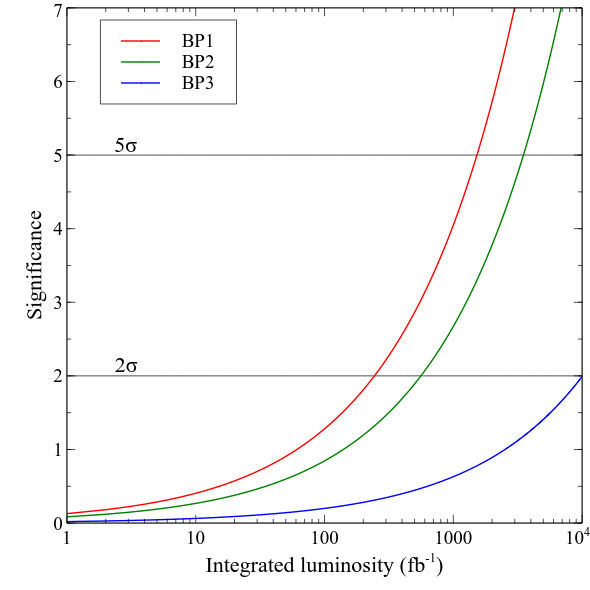}}
\caption{Variation of the statistical significance with the integrated luminosity.}
\label{f:signi}
\end{figure}

Table \ref{tab:BDT_yield} shows that the $j j +\met$ background can be hugely reduced and even eliminated through a multivariate analysis. The largest contribution comes from $\tau^+ \tau^- b \bar{b} + \met$. However the corresponding number of events reduces with the improvement of background rejection as $M_A$ increases. Fig.\ref{f:signi} displays the variation of the statistical significance \emph{w.r.t.} the integrated luminosity for the BPs. It is seen that a 5$\sigma$ discovery of a pseudoscalar of mass $\simeq$ 200 GeV (BP1) is possible at around 1500 fb$^{-1}$ integrated luminosity. BP1 is thus clearly within the discovery reach of the high luminosity Large Hadron Collider (HL-LHC). BP2 requires $\simeq 3200$ fb$^{-1}$ for the same indicating that the maximum $M_A$ that can be discovered at 5$\sigma$ at the HL-LHC is somewhere between 200 and 250 GeV. And BP3 is beyond such a reach.

Therefore, the success of the present analysis lies in predicting a 5$\sigma$ observability for $M_A \gtrsim$ 200 GeV through a $\tau_h\tau_h + \met$ signal where the bulk of the missing transverse energy comes from an invisible scalar. In hindsight, $p p \to H A \to \tau^+ \tau^- \tau^+ \tau^-, \tau^+ \tau^- \mu^+ \mu^-$ can also be promising in for the scalar mass ranges of interest in this study. While such a signal warrants a separate investigation, it remains completely agnostic of the particle sector leading to the requisite $\Delta a_\mu$ even for the heavier $A$. Therefore, this signal was not analysed in the present study and can be taken up for further scrutiny in the near future.

\begin{figure}
\centering
\subfigure[]{
\includegraphics[scale=0.40]{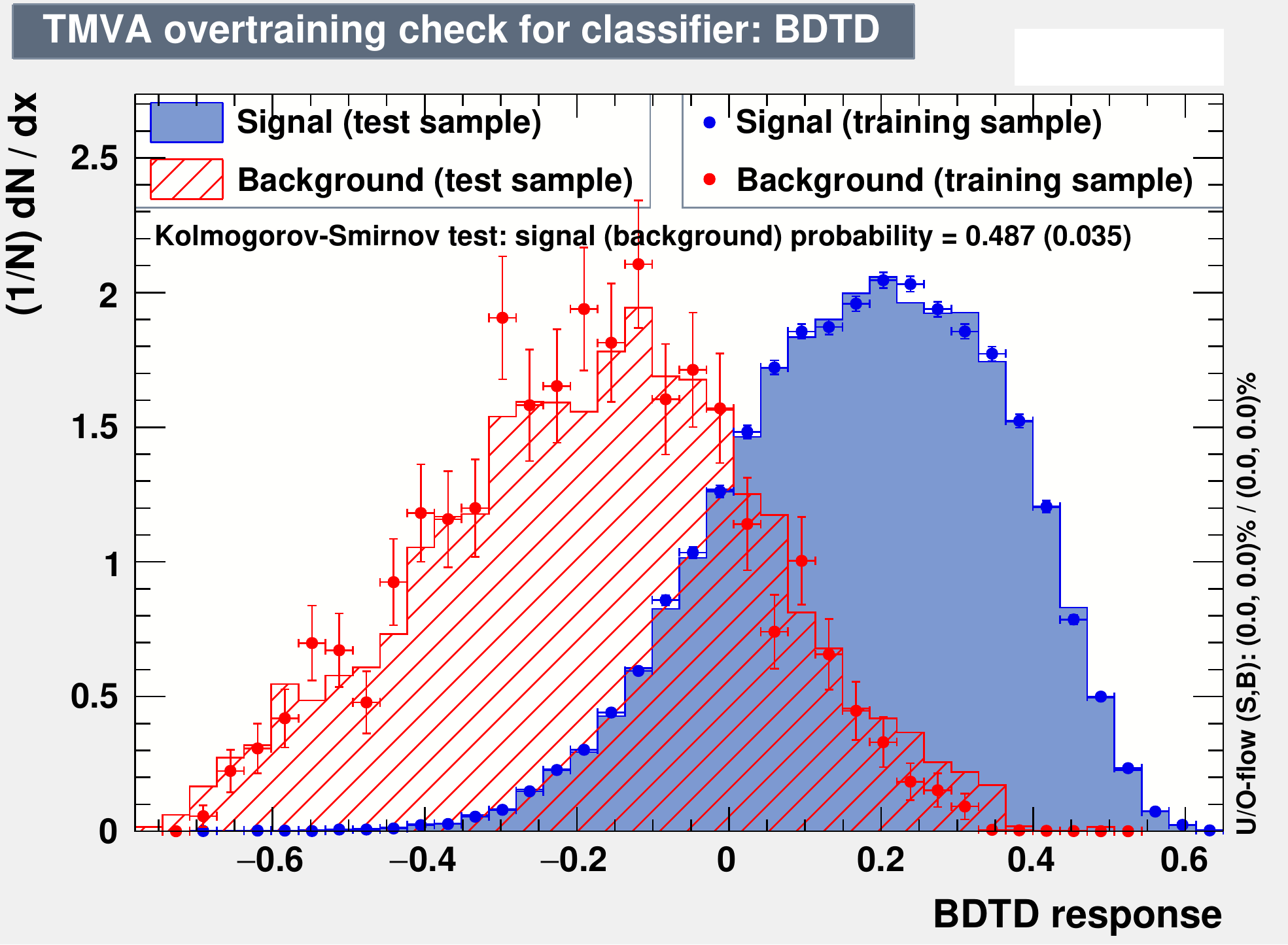}}
\subfigure[]{
\includegraphics[scale=0.40]{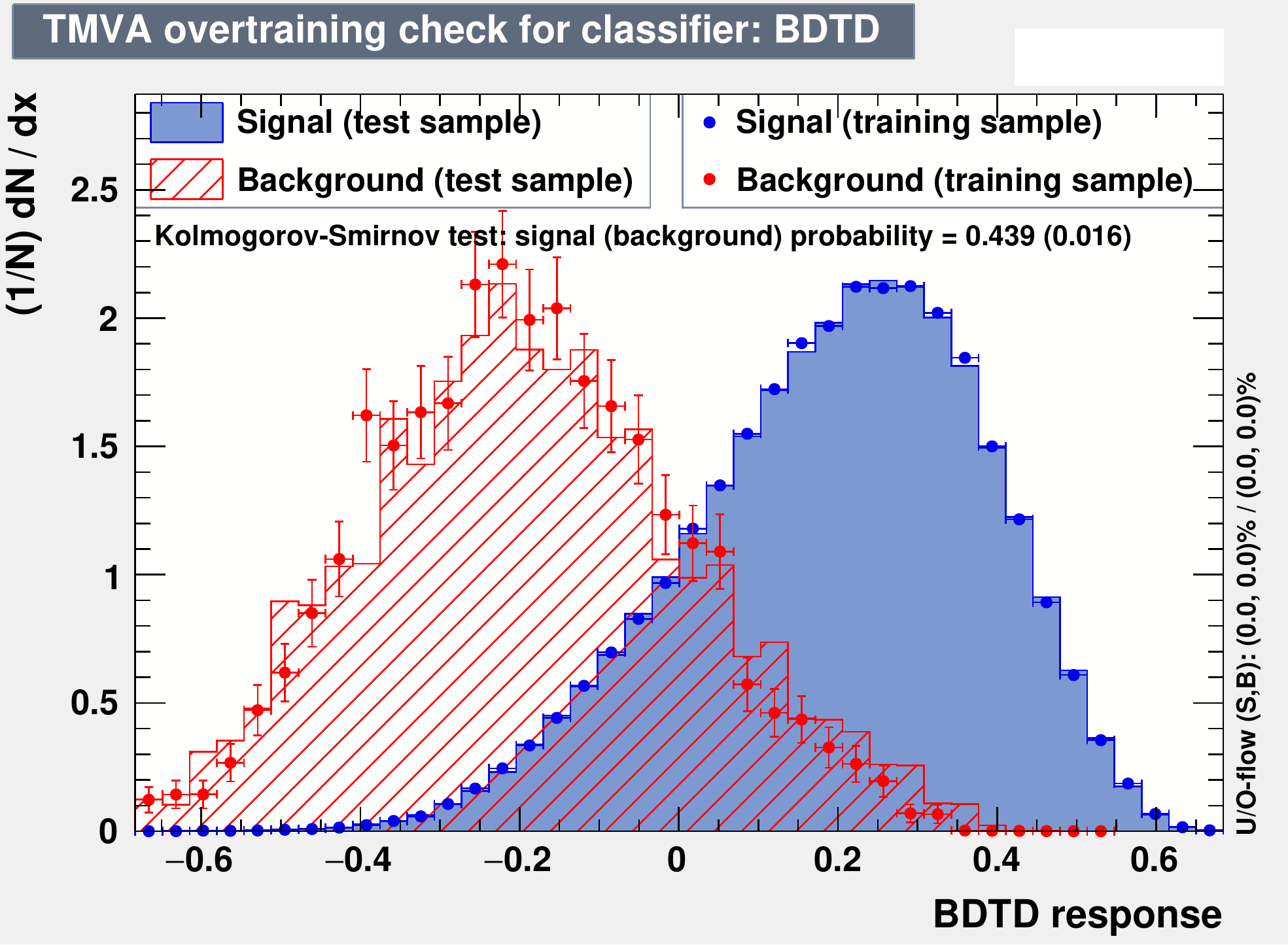}} \\
\subfigure[]{
\includegraphics[scale=0.40]{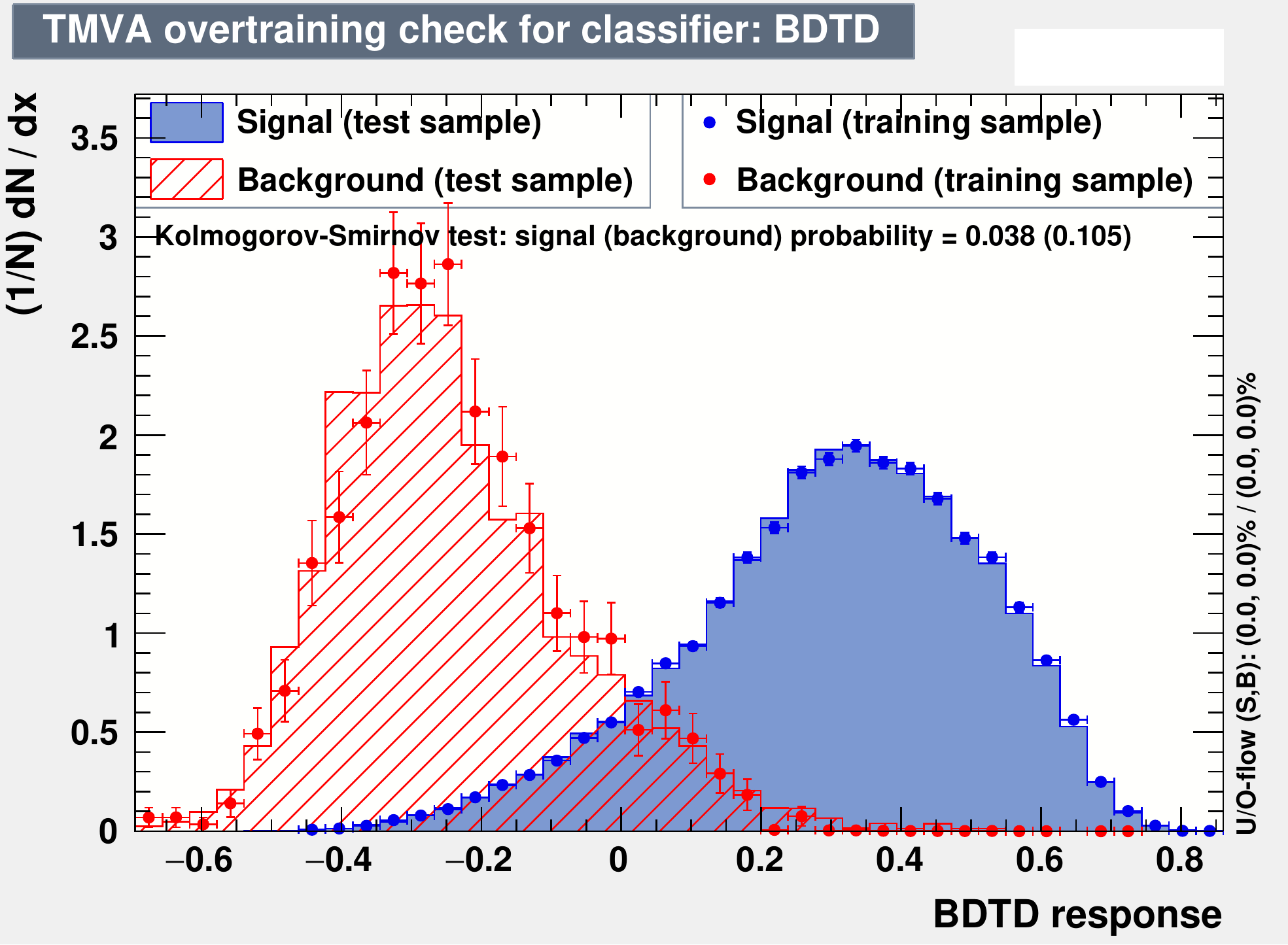}}
\caption{The KS-scores for the chosen lepton-specific benchmarks. The top-left, top-right and bottom panels correspond to BP1, BP2 and BP3 respectively.}
\label{f:KS}
\end{figure}

\section{Conclusions}\label{conclusions}

The Type-X 2HDM is long known to accommodate the observed excess in muon $g-2$ for high tan$\beta$ and low $M_A$. However, non-observation of $h \to A A$ at the LHC seriously limits the parameter region. We have augmented the Type-X 2HDM by an additional inert doublet in this work that is endowed with an additional $\mathbb{Z}_2$ symmetry. We have taken into account all constraints that are mandated by such a construct and thereafter compute the two-loop Barr-Zee contributions from the inert scalars to $\Delta a_\mu$. With these additional contributions, we demonstrate that a pesudoscalar mass as large as $M_A \sim$ 850 GeV becomes compatible with the observed $\Delta a_\mu$. The otherwise constrained parameter region in the 
$M_A$-tan$\beta$ plane obtained in case of the standalone 2HDM thus expands
to include much higher $M_A$. 

We have probed the scenario at the 14 TeV LHC through the signal 
$p p \to \eta_R \eta_A \to \eta_R \eta_R A \to \tau^+ \tau^- + \met$. We have considered a fully hadronic $\tau$ pair. 
Since the $\tau$-jets in this case originate from the \emph{heavier} $A$, certain kinematic features such as $p_T$ of the leading $\tau$-jet and the invariant and transverse masses of the pair are different compared to the SM and even the pure Type-X 2HDM. In addition, involvement of the inert scalar in the final state modifies the $\met$ spectrum too. We have exploited such a modified kinematics through a multivariate analysis of the signal and backgrounds using the BDTD algorithm. We subsequently predict a 5$\sigma$ discovery for $M_A \gtrsim$ 200 GeV at the HL-LHC.

\acknowledgements NC acknowledges support from DST, India, under
Grant Number IFA19-PH237 (INSPIRE Faculty Award). NC thanks Indrani Chakraborty for fruitful discussions on TMVA.

\section{Appendix}
\subsection{Trilinear couplings}\label{appen_A}
\besub
\bea
\l_{h \eta^+ \eta^-} &=& 
(\s_1 c_{\a+\b} - \nu_1 s_\a c_\b + \nu_2 c_\a s_\b), \\
\l_{H \eta^+ \eta^-} &=& 
(\s_1 s_{\a+\b} + \nu_1 c_\a c_\b + \nu_2 s_\a s_\b), \\
\l_{A \eta_R \eta_I} &=& 2 
(\s_3 c_{2\b} + (-\k_1 + \k_2)c_\b s_\b), \\
\l_{H^+ \eta^- \eta_R} &=& \big((\s_2 + \s_3)c_{2\b} + (-\k_1 + \k_2 - \omega_1 + \omega_2)s_\b c_\b \big), \\
\l_{H^+ \eta^- \eta_I} &=& \big((\s_2 - \s_3)c_{2\b} + (\k_1 - \k_2 - \omega_1 + \omega_2)s_\b c_\b \big).
\eea
\eesub

\subsection{Two loop functions}\label{appen_B}
\besub
\bea
\mathcal{F}^{(1)} (z) &=& \frac{z}{2} \int_{0}^{1} dx \frac{2x(1-x)-1}{z-x(1-x)} ~{\rm ln} \left(\frac{z}{x(1-x)}\right), \\
\tilde{\mathcal{F}}^{(1)} (z) &=& \frac{z}{2} \int_{0}^{1} dx \frac{1}{z-x(1-x)} ~{\rm ln} \left(\frac{z}{x(1-x)}\right), \\
\mathcal{F}^{(2)} (z) &=& \frac{1}{2} \int_{0}^{1} dx \frac{x(1-x)}{z-x(1-x)} ~{\rm ln} \left(\frac{z}{x(1-x)}\right), \\
\mathcal{G}(z^a,z^b,x) &=& \frac{{\rm ln} \left(\frac{z^a x + z^b (1-x)}{x(1-x)}\right)}{x(1-x) - z^a x - z^b (1-x)}.
\eea
\eesub

\bibliography{ref_2plus1} 

\begin{thebibliography}{62}%
\makeatletter
\providecommand \@ifxundefined [1]{%
 \@ifx{#1\undefined}
}%
\providecommand \@ifnum [1]{%
 \ifnum #1\expandafter \@firstoftwo
 \else \expandafter \@secondoftwo
 \fi
}%
\providecommand \@ifx [1]{%
 \ifx #1\expandafter \@firstoftwo
 \else \expandafter \@secondoftwo
 \fi
}%
\providecommand \natexlab [1]{#1}%
\providecommand \enquote  [1]{``#1''}%
\providecommand \bibnamefont  [1]{#1}%
\providecommand \bibfnamefont [1]{#1}%
\providecommand \citenamefont [1]{#1}%
\providecommand \href@noop [0]{\@secondoftwo}%
\providecommand \href [0]{\begingroup \@sanitize@url \@href}%
\providecommand \@href[1]{\@@startlink{#1}\@@href}%
\providecommand \@@href[1]{\endgroup#1\@@endlink}%
\providecommand \@sanitize@url [0]{\catcode `\\12\catcode `\$12\catcode
  `\&12\catcode `\#12\catcode `\^12\catcode `\_12\catcode `\%12\relax}%
\providecommand \@@startlink[1]{}%
\providecommand \@@endlink[0]{}%
\providecommand \url  [0]{\begingroup\@sanitize@url \@url }%
\providecommand \@url [1]{\endgroup\@href {#1}{\urlprefix }}%
\providecommand \urlprefix  [0]{URL }%
\providecommand \Eprint [0]{\href }%
\providecommand \doibase [0]{http://dx.doi.org/}%
\providecommand \selectlanguage [0]{\@gobble}%
\providecommand \bibinfo  [0]{\@secondoftwo}%
\providecommand \bibfield  [0]{\@secondoftwo}%
\providecommand \translation [1]{[#1]}%
\providecommand \BibitemOpen [0]{}%
\providecommand \bibitemStop [0]{}%
\providecommand \bibitemNoStop [0]{.\EOS\space}%
\providecommand \EOS [0]{\spacefactor3000\relax}%
\providecommand \BibitemShut  [1]{\csname bibitem#1\endcsname}%
\let\auto@bib@innerbib\@empty
%</preamble>
\bibitem [{\citenamefont {Blum}\ \emph {et~al.}(2013)\citenamefont {Blum},
  \citenamefont {Denig}, \citenamefont {Logashenko}, \citenamefont {de~Rafael},
  \citenamefont {Roberts}, \citenamefont {Teubner},\ and\ \citenamefont
  {Venanzoni}}]{Blum:2013xva}%
  \BibitemOpen
  \bibfield  {author} {\bibinfo {author} {\bibfnamefont {T.}~\bibnamefont
  {Blum}}, \bibinfo {author} {\bibfnamefont {A.}~\bibnamefont {Denig}},
  \bibinfo {author} {\bibfnamefont {I.}~\bibnamefont {Logashenko}}, \bibinfo
  {author} {\bibfnamefont {E.}~\bibnamefont {de~Rafael}}, \bibinfo {author}
  {\bibfnamefont {B.~L.}\ \bibnamefont {Roberts}}, \bibinfo {author}
  {\bibfnamefont {T.}~\bibnamefont {Teubner}}, \ and\ \bibinfo {author}
  {\bibfnamefont {G.}~\bibnamefont {Venanzoni}},\ }\href@noop {} {\  (\bibinfo
  {year} {2013})},\ \Eprint {http://arxiv.org/abs/1311.2198} {arXiv:1311.2198
  [hep-ph]} \BibitemShut {NoStop}%
\bibitem [{\citenamefont {Blum}\ \emph {et~al.}(2018)\citenamefont {Blum},
  \citenamefont {Boyle}, \citenamefont {G\"ulpers}, \citenamefont {Izubuchi},
  \citenamefont {Jin}, \citenamefont {Jung}, \citenamefont {J\"uttner},
  \citenamefont {Lehner}, \citenamefont {Portelli},\ and\ \citenamefont
  {Tsang}}]{RBC:2018dos}%
  \BibitemOpen
  \bibfield  {author} {\bibinfo {author} {\bibfnamefont {T.}~\bibnamefont
  {Blum}}, \bibinfo {author} {\bibfnamefont {P.~A.}\ \bibnamefont {Boyle}},
  \bibinfo {author} {\bibfnamefont {V.}~\bibnamefont {G\"ulpers}}, \bibinfo
  {author} {\bibfnamefont {T.}~\bibnamefont {Izubuchi}}, \bibinfo {author}
  {\bibfnamefont {L.}~\bibnamefont {Jin}}, \bibinfo {author} {\bibfnamefont
  {C.}~\bibnamefont {Jung}}, \bibinfo {author} {\bibfnamefont {A.}~\bibnamefont
  {J\"uttner}}, \bibinfo {author} {\bibfnamefont {C.}~\bibnamefont {Lehner}},
  \bibinfo {author} {\bibfnamefont {A.}~\bibnamefont {Portelli}}, \ and\
  \bibinfo {author} {\bibfnamefont {J.~T.}\ \bibnamefont {Tsang}} (\bibinfo
  {collaboration} {RBC, UKQCD}),\ }\href {\doibase
  10.1103/PhysRevLett.121.022003} {\bibfield  {journal} {\bibinfo  {journal}
  {Phys. Rev. Lett.}\ }\textbf {\bibinfo {volume} {121}},\ \bibinfo {pages}
  {022003} (\bibinfo {year} {2018})},\ \Eprint
  {http://arxiv.org/abs/1801.07224} {arXiv:1801.07224 [hep-lat]} \BibitemShut
  {NoStop}%
\bibitem [{\citenamefont {Keshavarzi}\ \emph {et~al.}(2018)\citenamefont
  {Keshavarzi}, \citenamefont {Nomura},\ and\ \citenamefont
  {Teubner}}]{Keshavarzi:2018mgv}%
  \BibitemOpen
  \bibfield  {author} {\bibinfo {author} {\bibfnamefont {A.}~\bibnamefont
  {Keshavarzi}}, \bibinfo {author} {\bibfnamefont {D.}~\bibnamefont {Nomura}},
  \ and\ \bibinfo {author} {\bibfnamefont {T.}~\bibnamefont {Teubner}},\ }\href
  {\doibase 10.1103/PhysRevD.97.114025} {\bibfield  {journal} {\bibinfo
  {journal} {Phys. Rev. D}\ }\textbf {\bibinfo {volume} {97}},\ \bibinfo
  {pages} {114025} (\bibinfo {year} {2018})},\ \Eprint
  {http://arxiv.org/abs/1802.02995} {arXiv:1802.02995 [hep-ph]} \BibitemShut
  {NoStop}%
\bibitem [{\citenamefont {Davier}\ \emph {et~al.}(2020)\citenamefont {Davier},
  \citenamefont {Hoecker}, \citenamefont {Malaescu},\ and\ \citenamefont
  {Zhang}}]{Davier:2019can}%
  \BibitemOpen
  \bibfield  {author} {\bibinfo {author} {\bibfnamefont {M.}~\bibnamefont
  {Davier}}, \bibinfo {author} {\bibfnamefont {A.}~\bibnamefont {Hoecker}},
  \bibinfo {author} {\bibfnamefont {B.}~\bibnamefont {Malaescu}}, \ and\
  \bibinfo {author} {\bibfnamefont {Z.}~\bibnamefont {Zhang}},\ }\href
  {\doibase 10.1140/epjc/s10052-020-7792-2} {\bibfield  {journal} {\bibinfo
  {journal} {Eur. Phys. J. C}\ }\textbf {\bibinfo {volume} {80}},\ \bibinfo
  {pages} {241} (\bibinfo {year} {2020})},\ \bibinfo {note} {[Erratum:
  Eur.Phys.J.C 80, 410 (2020)]},\ \Eprint {http://arxiv.org/abs/1908.00921}
  {arXiv:1908.00921 [hep-ph]} \BibitemShut {NoStop}%
\bibitem [{\citenamefont {Aoyama}\ \emph {et~al.}(2020)\citenamefont {Aoyama}
  \emph {et~al.}}]{Aoyama:2020ynm}%
  \BibitemOpen
  \bibfield  {author} {\bibinfo {author} {\bibfnamefont {T.}~\bibnamefont
  {Aoyama}} \emph {et~al.},\ }\href {\doibase 10.1016/j.physrep.2020.07.006}
  {\bibfield  {journal} {\bibinfo  {journal} {Phys. Rept.}\ }\textbf {\bibinfo
  {volume} {887}},\ \bibinfo {pages} {1} (\bibinfo {year} {2020})},\ \Eprint
  {http://arxiv.org/abs/2006.04822} {arXiv:2006.04822 [hep-ph]} \BibitemShut
  {NoStop}%
\bibitem [{\citenamefont {Colangelo}\ \emph {et~al.}(2019)\citenamefont
  {Colangelo}, \citenamefont {Hoferichter},\ and\ \citenamefont
  {Stoffer}}]{Colangelo:2018mtw}%
  \BibitemOpen
  \bibfield  {author} {\bibinfo {author} {\bibfnamefont {G.}~\bibnamefont
  {Colangelo}}, \bibinfo {author} {\bibfnamefont {M.}~\bibnamefont
  {Hoferichter}}, \ and\ \bibinfo {author} {\bibfnamefont {P.}~\bibnamefont
  {Stoffer}},\ }\href {\doibase 10.1007/JHEP02(2019)006} {\bibfield  {journal}
  {\bibinfo  {journal} {JHEP}\ }\textbf {\bibinfo {volume} {02}},\ \bibinfo
  {pages} {006} (\bibinfo {year} {2019})},\ \Eprint
  {http://arxiv.org/abs/1810.00007} {arXiv:1810.00007 [hep-ph]} \BibitemShut
  {NoStop}%
\bibitem [{\citenamefont {Hoferichter}\ \emph {et~al.}(2019)\citenamefont
  {Hoferichter}, \citenamefont {Hoid},\ and\ \citenamefont
  {Kubis}}]{Hoferichter:2019mqg}%
  \BibitemOpen
  \bibfield  {author} {\bibinfo {author} {\bibfnamefont {M.}~\bibnamefont
  {Hoferichter}}, \bibinfo {author} {\bibfnamefont {B.-L.}\ \bibnamefont
  {Hoid}}, \ and\ \bibinfo {author} {\bibfnamefont {B.}~\bibnamefont {Kubis}},\
  }\href {\doibase 10.1007/JHEP08(2019)137} {\bibfield  {journal} {\bibinfo
  {journal} {JHEP}\ }\textbf {\bibinfo {volume} {08}},\ \bibinfo {pages} {137}
  (\bibinfo {year} {2019})},\ \Eprint {http://arxiv.org/abs/1907.01556}
  {arXiv:1907.01556 [hep-ph]} \BibitemShut {NoStop}%
\bibitem [{\citenamefont {Melnikov}\ and\ \citenamefont
  {Vainshtein}(2004)}]{Melnikov:2003xd}%
  \BibitemOpen
  \bibfield  {author} {\bibinfo {author} {\bibfnamefont {K.}~\bibnamefont
  {Melnikov}}\ and\ \bibinfo {author} {\bibfnamefont {A.}~\bibnamefont
  {Vainshtein}},\ }\href {\doibase 10.1103/PhysRevD.70.113006} {\bibfield
  {journal} {\bibinfo  {journal} {Phys. Rev. D}\ }\textbf {\bibinfo {volume}
  {70}},\ \bibinfo {pages} {113006} (\bibinfo {year} {2004})},\ \Eprint
  {http://arxiv.org/abs/hep-ph/0312226} {arXiv:hep-ph/0312226} \BibitemShut
  {NoStop}%
\bibitem [{\citenamefont {Hoferichter}\ \emph {et~al.}(2018)\citenamefont
  {Hoferichter}, \citenamefont {Hoid}, \citenamefont {Kubis}, \citenamefont
  {Leupold},\ and\ \citenamefont {Schneider}}]{Hoferichter:2018kwz}%
  \BibitemOpen
  \bibfield  {author} {\bibinfo {author} {\bibfnamefont {M.}~\bibnamefont
  {Hoferichter}}, \bibinfo {author} {\bibfnamefont {B.-L.}\ \bibnamefont
  {Hoid}}, \bibinfo {author} {\bibfnamefont {B.}~\bibnamefont {Kubis}},
  \bibinfo {author} {\bibfnamefont {S.}~\bibnamefont {Leupold}}, \ and\
  \bibinfo {author} {\bibfnamefont {S.~P.}\ \bibnamefont {Schneider}},\ }\href
  {\doibase 10.1007/JHEP10(2018)141} {\bibfield  {journal} {\bibinfo  {journal}
  {JHEP}\ }\textbf {\bibinfo {volume} {10}},\ \bibinfo {pages} {141} (\bibinfo
  {year} {2018})},\ \Eprint {http://arxiv.org/abs/1808.04823} {arXiv:1808.04823
  [hep-ph]} \BibitemShut {NoStop}%
\bibitem [{\citenamefont {Blum}\ \emph {et~al.}(2020)\citenamefont {Blum},
  \citenamefont {Christ}, \citenamefont {Hayakawa}, \citenamefont {Izubuchi},
  \citenamefont {Jin}, \citenamefont {Jung},\ and\ \citenamefont
  {Lehner}}]{Blum:2019ugy}%
  \BibitemOpen
  \bibfield  {author} {\bibinfo {author} {\bibfnamefont {T.}~\bibnamefont
  {Blum}}, \bibinfo {author} {\bibfnamefont {N.}~\bibnamefont {Christ}},
  \bibinfo {author} {\bibfnamefont {M.}~\bibnamefont {Hayakawa}}, \bibinfo
  {author} {\bibfnamefont {T.}~\bibnamefont {Izubuchi}}, \bibinfo {author}
  {\bibfnamefont {L.}~\bibnamefont {Jin}}, \bibinfo {author} {\bibfnamefont
  {C.}~\bibnamefont {Jung}}, \ and\ \bibinfo {author} {\bibfnamefont
  {C.}~\bibnamefont {Lehner}},\ }\href {\doibase
  10.1103/PhysRevLett.124.132002} {\bibfield  {journal} {\bibinfo  {journal}
  {Phys. Rev. Lett.}\ }\textbf {\bibinfo {volume} {124}},\ \bibinfo {pages}
  {132002} (\bibinfo {year} {2020})},\ \Eprint
  {http://arxiv.org/abs/1911.08123} {arXiv:1911.08123 [hep-lat]} \BibitemShut
  {NoStop}%
\bibitem [{\citenamefont {Zyla}\ \emph {et~al.}(2020)\citenamefont {Zyla} \emph
  {et~al.}}]{ParticleDataGroup:2020ssz}%
  \BibitemOpen
  \bibfield  {author} {\bibinfo {author} {\bibfnamefont {P.~A.}\ \bibnamefont
  {Zyla}} \emph {et~al.} (\bibinfo {collaboration} {Particle Data Group}),\
  }\href {\doibase 10.1093/ptep/ptaa104} {\bibfield  {journal} {\bibinfo
  {journal} {PTEP}\ }\textbf {\bibinfo {volume} {2020}},\ \bibinfo {pages}
  {083C01} (\bibinfo {year} {2020})}\BibitemShut {NoStop}%
\bibitem [{\citenamefont {Bennett}\ \emph {et~al.}(2006)\citenamefont {Bennett}
  \emph {et~al.}}]{Muong-2:2006rrc}%
  \BibitemOpen
  \bibfield  {author} {\bibinfo {author} {\bibfnamefont {G.~W.}\ \bibnamefont
  {Bennett}} \emph {et~al.} (\bibinfo {collaboration} {Muon g-2}),\ }\href
  {\doibase 10.1103/PhysRevD.73.072003} {\bibfield  {journal} {\bibinfo
  {journal} {Phys. Rev. D}\ }\textbf {\bibinfo {volume} {73}},\ \bibinfo
  {pages} {072003} (\bibinfo {year} {2006})},\ \Eprint
  {http://arxiv.org/abs/hep-ex/0602035} {arXiv:hep-ex/0602035} \BibitemShut
  {NoStop}%
\bibitem [{\citenamefont {Abi}\ \emph {et~al.}(2021)\citenamefont {Abi} \emph
  {et~al.}}]{Muong-2:2021ojo}%
  \BibitemOpen
  \bibfield  {author} {\bibinfo {author} {\bibfnamefont {B.}~\bibnamefont
  {Abi}} \emph {et~al.} (\bibinfo {collaboration} {Muon g-2}),\ }\href
  {\doibase 10.1103/PhysRevLett.126.141801} {\bibfield  {journal} {\bibinfo
  {journal} {Phys. Rev. Lett.}\ }\textbf {\bibinfo {volume} {126}},\ \bibinfo
  {pages} {141801} (\bibinfo {year} {2021})},\ \Eprint
  {http://arxiv.org/abs/2104.03281} {arXiv:2104.03281 [hep-ex]} \BibitemShut
  {NoStop}%
\bibitem [{\citenamefont {Albahri}\ \emph {et~al.}(2021)\citenamefont {Albahri}
  \emph {et~al.}}]{Muong-2:2021vma}%
  \BibitemOpen
  \bibfield  {author} {\bibinfo {author} {\bibfnamefont {T.}~\bibnamefont
  {Albahri}} \emph {et~al.} (\bibinfo {collaboration} {Muon g-2}),\ }\href
  {\doibase 10.1103/PhysRevD.103.072002} {\bibfield  {journal} {\bibinfo
  {journal} {Phys. Rev. D}\ }\textbf {\bibinfo {volume} {103}},\ \bibinfo
  {pages} {072002} (\bibinfo {year} {2021})},\ \Eprint
  {http://arxiv.org/abs/2104.03247} {arXiv:2104.03247 [hep-ex]} \BibitemShut
  {NoStop}%
\bibitem [{\citenamefont {Deshpande}\ and\ \citenamefont
  {Ma}(1978)}]{Deshpande:1977rw}%
  \BibitemOpen
  \bibfield  {author} {\bibinfo {author} {\bibfnamefont {N.~G.}\ \bibnamefont
  {Deshpande}}\ and\ \bibinfo {author} {\bibfnamefont {E.}~\bibnamefont {Ma}},\
  }\href {\doibase 10.1103/PhysRevD.18.2574} {\bibfield  {journal} {\bibinfo
  {journal} {Phys. Rev. D}\ }\textbf {\bibinfo {volume} {18}},\ \bibinfo
  {pages} {2574} (\bibinfo {year} {1978})}\BibitemShut {NoStop}%
\bibitem [{\citenamefont {Branco}\ \emph {et~al.}(2012)\citenamefont {Branco},
  \citenamefont {Ferreira}, \citenamefont {Lavoura}, \citenamefont {Rebelo},
  \citenamefont {Sher},\ and\ \citenamefont {Silva}}]{Branco:2011iw}%
  \BibitemOpen
  \bibfield  {author} {\bibinfo {author} {\bibfnamefont {G.~C.}\ \bibnamefont
  {Branco}}, \bibinfo {author} {\bibfnamefont {P.~M.}\ \bibnamefont
  {Ferreira}}, \bibinfo {author} {\bibfnamefont {L.}~\bibnamefont {Lavoura}},
  \bibinfo {author} {\bibfnamefont {M.~N.}\ \bibnamefont {Rebelo}}, \bibinfo
  {author} {\bibfnamefont {M.}~\bibnamefont {Sher}}, \ and\ \bibinfo {author}
  {\bibfnamefont {J.~P.}\ \bibnamefont {Silva}},\ }\href {\doibase
  10.1016/j.physrep.2012.02.002} {\bibfield  {journal} {\bibinfo  {journal}
  {Phys. Rept.}\ }\textbf {\bibinfo {volume} {516}},\ \bibinfo {pages} {1}
  (\bibinfo {year} {2012})},\ \Eprint {http://arxiv.org/abs/1106.0034}
  {arXiv:1106.0034 [hep-ph]} \BibitemShut {NoStop}%
\bibitem [{\citenamefont {Broggio}\ \emph {et~al.}(2014)\citenamefont
  {Broggio}, \citenamefont {Chun}, \citenamefont {Passera}, \citenamefont
  {Patel},\ and\ \citenamefont {Vempati}}]{Broggio:2014mna}%
  \BibitemOpen
  \bibfield  {author} {\bibinfo {author} {\bibfnamefont {A.}~\bibnamefont
  {Broggio}}, \bibinfo {author} {\bibfnamefont {E.~J.}\ \bibnamefont {Chun}},
  \bibinfo {author} {\bibfnamefont {M.}~\bibnamefont {Passera}}, \bibinfo
  {author} {\bibfnamefont {K.~M.}\ \bibnamefont {Patel}}, \ and\ \bibinfo
  {author} {\bibfnamefont {S.~K.}\ \bibnamefont {Vempati}},\ }\href {\doibase
  10.1007/JHEP11(2014)058} {\bibfield  {journal} {\bibinfo  {journal} {JHEP}\
  }\textbf {\bibinfo {volume} {11}},\ \bibinfo {pages} {058} (\bibinfo {year}
  {2014})},\ \Eprint {http://arxiv.org/abs/1409.3199} {arXiv:1409.3199
  [hep-ph]} \BibitemShut {NoStop}%
\bibitem [{\citenamefont {Cao}\ \emph {et~al.}(2009)\citenamefont {Cao},
  \citenamefont {Wan}, \citenamefont {Wu},\ and\ \citenamefont
  {Yang}}]{Cao:2009as}%
  \BibitemOpen
  \bibfield  {author} {\bibinfo {author} {\bibfnamefont {J.}~\bibnamefont
  {Cao}}, \bibinfo {author} {\bibfnamefont {P.}~\bibnamefont {Wan}}, \bibinfo
  {author} {\bibfnamefont {L.}~\bibnamefont {Wu}}, \ and\ \bibinfo {author}
  {\bibfnamefont {J.~M.}\ \bibnamefont {Yang}},\ }\href {\doibase
  10.1103/PhysRevD.80.071701} {\bibfield  {journal} {\bibinfo  {journal} {Phys.
  Rev. D}\ }\textbf {\bibinfo {volume} {80}},\ \bibinfo {pages} {071701}
  (\bibinfo {year} {2009})},\ \Eprint {http://arxiv.org/abs/0909.5148}
  {arXiv:0909.5148 [hep-ph]} \BibitemShut {NoStop}%
\bibitem [{\citenamefont {Wang}\ and\ \citenamefont
  {Han}(2015)}]{Wang:2014sda}%
  \BibitemOpen
  \bibfield  {author} {\bibinfo {author} {\bibfnamefont {L.}~\bibnamefont
  {Wang}}\ and\ \bibinfo {author} {\bibfnamefont {X.-F.}\ \bibnamefont {Han}},\
  }\href {\doibase 10.1007/JHEP05(2015)039} {\bibfield  {journal} {\bibinfo
  {journal} {JHEP}\ }\textbf {\bibinfo {volume} {05}},\ \bibinfo {pages} {039}
  (\bibinfo {year} {2015})},\ \Eprint {http://arxiv.org/abs/1412.4874}
  {arXiv:1412.4874 [hep-ph]} \BibitemShut {NoStop}%
\bibitem [{\citenamefont {Ilisie}(2015)}]{Ilisie:2015tra}%
  \BibitemOpen
  \bibfield  {author} {\bibinfo {author} {\bibfnamefont {V.}~\bibnamefont
  {Ilisie}},\ }\href {\doibase 10.1007/JHEP04(2015)077} {\bibfield  {journal}
  {\bibinfo  {journal} {JHEP}\ }\textbf {\bibinfo {volume} {04}},\ \bibinfo
  {pages} {077} (\bibinfo {year} {2015})},\ \Eprint
  {http://arxiv.org/abs/1502.04199} {arXiv:1502.04199 [hep-ph]} \BibitemShut
  {NoStop}%
\bibitem [{\citenamefont {Abe}\ \emph {et~al.}(2015)\citenamefont {Abe},
  \citenamefont {Sato},\ and\ \citenamefont {Yagyu}}]{Abe:2015oca}%
  \BibitemOpen
  \bibfield  {author} {\bibinfo {author} {\bibfnamefont {T.}~\bibnamefont
  {Abe}}, \bibinfo {author} {\bibfnamefont {R.}~\bibnamefont {Sato}}, \ and\
  \bibinfo {author} {\bibfnamefont {K.}~\bibnamefont {Yagyu}},\ }\href
  {\doibase 10.1007/JHEP07(2015)064} {\bibfield  {journal} {\bibinfo  {journal}
  {JHEP}\ }\textbf {\bibinfo {volume} {07}},\ \bibinfo {pages} {064} (\bibinfo
  {year} {2015})},\ \Eprint {http://arxiv.org/abs/1504.07059} {arXiv:1504.07059
  [hep-ph]} \BibitemShut {NoStop}%
\bibitem [{\citenamefont {Chun}\ and\ \citenamefont
  {Kim}(2016)}]{Chun:2016hzs}%
  \BibitemOpen
  \bibfield  {author} {\bibinfo {author} {\bibfnamefont {E.~J.}\ \bibnamefont
  {Chun}}\ and\ \bibinfo {author} {\bibfnamefont {J.}~\bibnamefont {Kim}},\
  }\href {\doibase 10.1007/JHEP07(2016)110} {\bibfield  {journal} {\bibinfo
  {journal} {JHEP}\ }\textbf {\bibinfo {volume} {07}},\ \bibinfo {pages} {110}
  (\bibinfo {year} {2016})},\ \Eprint {http://arxiv.org/abs/1605.06298}
  {arXiv:1605.06298 [hep-ph]} \BibitemShut {NoStop}%
\bibitem [{\citenamefont {Cherchiglia}\ \emph {et~al.}(2017)\citenamefont
  {Cherchiglia}, \citenamefont {Kneschke}, \citenamefont {St\"ockinger},\ and\
  \citenamefont {St\"ockinger-Kim}}]{Cherchiglia:2016eui}%
  \BibitemOpen
  \bibfield  {author} {\bibinfo {author} {\bibfnamefont {A.}~\bibnamefont
  {Cherchiglia}}, \bibinfo {author} {\bibfnamefont {P.}~\bibnamefont
  {Kneschke}}, \bibinfo {author} {\bibfnamefont {D.}~\bibnamefont
  {St\"ockinger}}, \ and\ \bibinfo {author} {\bibfnamefont {H.}~\bibnamefont
  {St\"ockinger-Kim}},\ }\href {\doibase 10.1007/JHEP10(2021)242} {\bibfield
  {journal} {\bibinfo  {journal} {JHEP}\ }\textbf {\bibinfo {volume} {01}},\
  \bibinfo {pages} {007} (\bibinfo {year} {2017})},\ \bibinfo {note} {[Erratum:
  JHEP 10, 242 (2021)]},\ \Eprint {http://arxiv.org/abs/1607.06292}
  {arXiv:1607.06292 [hep-ph]} \BibitemShut {NoStop}%
\bibitem [{\citenamefont {Dey}\ \emph {et~al.}(2021)\citenamefont {Dey},
  \citenamefont {Lahiri},\ and\ \citenamefont {Mukhopadhyaya}}]{Dey:2021pyn}%
  \BibitemOpen
  \bibfield  {author} {\bibinfo {author} {\bibfnamefont {A.}~\bibnamefont
  {Dey}}, \bibinfo {author} {\bibfnamefont {J.}~\bibnamefont {Lahiri}}, \ and\
  \bibinfo {author} {\bibfnamefont {B.}~\bibnamefont {Mukhopadhyaya}},\
  }\href@noop {} {\  (\bibinfo {year} {2021})},\ \Eprint
  {http://arxiv.org/abs/2106.01449} {arXiv:2106.01449 [hep-ph]} \BibitemShut
  {NoStop}%
\bibitem [{\citenamefont {Han}\ \emph {et~al.}(2019)\citenamefont {Han},
  \citenamefont {Li}, \citenamefont {Wang},\ and\ \citenamefont
  {Zhang}}]{Han:2018znu}%
  \BibitemOpen
  \bibfield  {author} {\bibinfo {author} {\bibfnamefont {X.-F.}\ \bibnamefont
  {Han}}, \bibinfo {author} {\bibfnamefont {T.}~\bibnamefont {Li}}, \bibinfo
  {author} {\bibfnamefont {L.}~\bibnamefont {Wang}}, \ and\ \bibinfo {author}
  {\bibfnamefont {Y.}~\bibnamefont {Zhang}},\ }\href {\doibase
  10.1103/PhysRevD.99.095034} {\bibfield  {journal} {\bibinfo  {journal} {Phys.
  Rev. D}\ }\textbf {\bibinfo {volume} {99}},\ \bibinfo {pages} {095034}
  (\bibinfo {year} {2019})},\ \Eprint {http://arxiv.org/abs/1812.02449}
  {arXiv:1812.02449 [hep-ph]} \BibitemShut {NoStop}%
\bibitem [{\citenamefont {Chowdhury}\ and\ \citenamefont
  {Eberhardt}(2018)}]{Chowdhury:2017aav}%
  \BibitemOpen
  \bibfield  {author} {\bibinfo {author} {\bibfnamefont {D.}~\bibnamefont
  {Chowdhury}}\ and\ \bibinfo {author} {\bibfnamefont {O.}~\bibnamefont
  {Eberhardt}},\ }\href {\doibase 10.1007/JHEP05(2018)161} {\bibfield
  {journal} {\bibinfo  {journal} {JHEP}\ }\textbf {\bibinfo {volume} {05}},\
  \bibinfo {pages} {161} (\bibinfo {year} {2018})},\ \Eprint
  {http://arxiv.org/abs/1711.02095} {arXiv:1711.02095 [hep-ph]} \BibitemShut
  {NoStop}%
\bibitem [{\citenamefont {Sirunyan}\ \emph
  {et~al.}(2018{\natexlab{a}})\citenamefont {Sirunyan} \emph
  {et~al.}}]{CMS:2018qvj}%
  \BibitemOpen
  \bibfield  {author} {\bibinfo {author} {\bibfnamefont {A.~M.}\ \bibnamefont
  {Sirunyan}} \emph {et~al.} (\bibinfo {collaboration} {CMS}),\ }\href
  {\doibase 10.1007/JHEP11(2018)018} {\bibfield  {journal} {\bibinfo  {journal}
  {JHEP}\ }\textbf {\bibinfo {volume} {11}},\ \bibinfo {pages} {018} (\bibinfo
  {year} {2018}{\natexlab{a}})},\ \Eprint {http://arxiv.org/abs/1805.04865}
  {arXiv:1805.04865 [hep-ex]} \BibitemShut {NoStop}%
\bibitem [{\citenamefont {Chun}\ \emph {et~al.}(2017)\citenamefont {Chun},
  \citenamefont {Dwivedi}, \citenamefont {Mondal},\ and\ \citenamefont
  {Mukhopadhyaya}}]{Chun:2017yob}%
  \BibitemOpen
  \bibfield  {author} {\bibinfo {author} {\bibfnamefont {E.~J.}\ \bibnamefont
  {Chun}}, \bibinfo {author} {\bibfnamefont {S.}~\bibnamefont {Dwivedi}},
  \bibinfo {author} {\bibfnamefont {T.}~\bibnamefont {Mondal}}, \ and\ \bibinfo
  {author} {\bibfnamefont {B.}~\bibnamefont {Mukhopadhyaya}},\ }\href {\doibase
  10.1016/j.physletb.2017.09.037} {\bibfield  {journal} {\bibinfo  {journal}
  {Phys. Lett. B}\ }\textbf {\bibinfo {volume} {774}},\ \bibinfo {pages} {20}
  (\bibinfo {year} {2017})},\ \Eprint {http://arxiv.org/abs/1707.07928}
  {arXiv:1707.07928 [hep-ph]} \BibitemShut {NoStop}%
\bibitem [{\citenamefont {Chun}\ \emph {et~al.}(2018)\citenamefont {Chun},
  \citenamefont {Dwivedi}, \citenamefont {Mondal}, \citenamefont
  {Mukhopadhyaya},\ and\ \citenamefont {Rai}}]{Chun:2018vsn}%
  \BibitemOpen
  \bibfield  {author} {\bibinfo {author} {\bibfnamefont {E.~J.}\ \bibnamefont
  {Chun}}, \bibinfo {author} {\bibfnamefont {S.}~\bibnamefont {Dwivedi}},
  \bibinfo {author} {\bibfnamefont {T.}~\bibnamefont {Mondal}}, \bibinfo
  {author} {\bibfnamefont {B.}~\bibnamefont {Mukhopadhyaya}}, \ and\ \bibinfo
  {author} {\bibfnamefont {S.~K.}\ \bibnamefont {Rai}},\ }\href {\doibase
  10.1103/PhysRevD.98.075008} {\bibfield  {journal} {\bibinfo  {journal} {Phys.
  Rev. D}\ }\textbf {\bibinfo {volume} {98}},\ \bibinfo {pages} {075008}
  (\bibinfo {year} {2018})},\ \Eprint {http://arxiv.org/abs/1807.05379}
  {arXiv:1807.05379 [hep-ph]} \BibitemShut {NoStop}%
\bibitem [{\citenamefont {Crivellin}\ \emph {et~al.}(2016)\citenamefont
  {Crivellin}, \citenamefont {Heeck},\ and\ \citenamefont
  {Stoffer}}]{Crivellin:2015hha}%
  \BibitemOpen
  \bibfield  {author} {\bibinfo {author} {\bibfnamefont {A.}~\bibnamefont
  {Crivellin}}, \bibinfo {author} {\bibfnamefont {J.}~\bibnamefont {Heeck}}, \
  and\ \bibinfo {author} {\bibfnamefont {P.}~\bibnamefont {Stoffer}},\ }\href
  {\doibase 10.1103/PhysRevLett.116.081801} {\bibfield  {journal} {\bibinfo
  {journal} {Phys. Rev. Lett.}\ }\textbf {\bibinfo {volume} {116}},\ \bibinfo
  {pages} {081801} (\bibinfo {year} {2016})},\ \Eprint
  {http://arxiv.org/abs/1507.07567} {arXiv:1507.07567 [hep-ph]} \BibitemShut
  {NoStop}%
\bibitem [{\citenamefont {Iguro}\ \emph {et~al.}(2019)\citenamefont {Iguro},
  \citenamefont {Omura},\ and\ \citenamefont {Takeuchi}}]{Iguro:2019sly}%
  \BibitemOpen
  \bibfield  {author} {\bibinfo {author} {\bibfnamefont {S.}~\bibnamefont
  {Iguro}}, \bibinfo {author} {\bibfnamefont {Y.}~\bibnamefont {Omura}}, \ and\
  \bibinfo {author} {\bibfnamefont {M.}~\bibnamefont {Takeuchi}},\ }\href
  {\doibase 10.1007/JHEP11(2019)130} {\bibfield  {journal} {\bibinfo  {journal}
  {JHEP}\ }\textbf {\bibinfo {volume} {11}},\ \bibinfo {pages} {130} (\bibinfo
  {year} {2019})},\ \Eprint {http://arxiv.org/abs/1907.09845} {arXiv:1907.09845
  [hep-ph]} \BibitemShut {NoStop}%
\bibitem [{\citenamefont {Jueid}\ \emph {et~al.}(2021)\citenamefont {Jueid},
  \citenamefont {Kim}, \citenamefont {Lee},\ and\ \citenamefont
  {Song}}]{Jueid:2021avn}%
  \BibitemOpen
  \bibfield  {author} {\bibinfo {author} {\bibfnamefont {A.}~\bibnamefont
  {Jueid}}, \bibinfo {author} {\bibfnamefont {J.}~\bibnamefont {Kim}}, \bibinfo
  {author} {\bibfnamefont {S.}~\bibnamefont {Lee}}, \ and\ \bibinfo {author}
  {\bibfnamefont {J.}~\bibnamefont {Song}},\ }\href {\doibase
  10.1103/PhysRevD.104.095008} {\bibfield  {journal} {\bibinfo  {journal}
  {Phys. Rev. D}\ }\textbf {\bibinfo {volume} {104}},\ \bibinfo {pages}
  {095008} (\bibinfo {year} {2021})},\ \Eprint
  {http://arxiv.org/abs/2104.10175} {arXiv:2104.10175 [hep-ph]} \BibitemShut
  {NoStop}%
\bibitem [{\citenamefont {Chen}\ \emph {et~al.}(2021)\citenamefont {Chen},
  \citenamefont {Wang},\ and\ \citenamefont {Yao}}]{Chen:2021rnl}%
  \BibitemOpen
  \bibfield  {author} {\bibinfo {author} {\bibfnamefont {N.}~\bibnamefont
  {Chen}}, \bibinfo {author} {\bibfnamefont {B.}~\bibnamefont {Wang}}, \ and\
  \bibinfo {author} {\bibfnamefont {C.-Y.}\ \bibnamefont {Yao}},\ }\href@noop
  {} {\  (\bibinfo {year} {2021})},\ \Eprint {http://arxiv.org/abs/2102.05619}
  {arXiv:2102.05619 [hep-ph]} \BibitemShut {NoStop}%
\bibitem [{\citenamefont {Wang}\ \emph {et~al.}(2019)\citenamefont {Wang},
  \citenamefont {Yang}, \citenamefont {Zhang},\ and\ \citenamefont
  {Zhang}}]{Wang:2018hnw}%
  \BibitemOpen
  \bibfield  {author} {\bibinfo {author} {\bibfnamefont {L.}~\bibnamefont
  {Wang}}, \bibinfo {author} {\bibfnamefont {J.~M.}\ \bibnamefont {Yang}},
  \bibinfo {author} {\bibfnamefont {M.}~\bibnamefont {Zhang}}, \ and\ \bibinfo
  {author} {\bibfnamefont {Y.}~\bibnamefont {Zhang}},\ }\href {\doibase
  10.1016/j.physletb.2018.11.045} {\bibfield  {journal} {\bibinfo  {journal}
  {Phys. Lett. B}\ }\textbf {\bibinfo {volume} {788}},\ \bibinfo {pages} {519}
  (\bibinfo {year} {2019})},\ \Eprint {http://arxiv.org/abs/1809.05857}
  {arXiv:1809.05857 [hep-ph]} \BibitemShut {NoStop}%
\bibitem [{\citenamefont {Frank}\ and\ \citenamefont
  {Saha}(2020)}]{Frank:2020smf}%
  \BibitemOpen
  \bibfield  {author} {\bibinfo {author} {\bibfnamefont {M.}~\bibnamefont
  {Frank}}\ and\ \bibinfo {author} {\bibfnamefont {I.}~\bibnamefont {Saha}},\
  }\href {\doibase 10.1103/PhysRevD.102.115034} {\bibfield  {journal} {\bibinfo
   {journal} {Phys. Rev. D}\ }\textbf {\bibinfo {volume} {102}},\ \bibinfo
  {pages} {115034} (\bibinfo {year} {2020})},\ \Eprint
  {http://arxiv.org/abs/2008.11909} {arXiv:2008.11909 [hep-ph]} \BibitemShut
  {NoStop}%
\bibitem [{\citenamefont {Chun}\ and\ \citenamefont
  {Mondal}(2020)}]{Chun:2020uzw}%
  \BibitemOpen
  \bibfield  {author} {\bibinfo {author} {\bibfnamefont {E.~J.}\ \bibnamefont
  {Chun}}\ and\ \bibinfo {author} {\bibfnamefont {T.}~\bibnamefont {Mondal}},\
  }\href {\doibase 10.1007/JHEP11(2020)077} {\bibfield  {journal} {\bibinfo
  {journal} {JHEP}\ }\textbf {\bibinfo {volume} {11}},\ \bibinfo {pages} {077}
  (\bibinfo {year} {2020})},\ \Eprint {http://arxiv.org/abs/2009.08314}
  {arXiv:2009.08314 [hep-ph]} \BibitemShut {NoStop}%
\bibitem [{\citenamefont {Moretti}\ and\ \citenamefont
  {Yagyu}(2015)}]{Moretti:2015cwa}%
  \BibitemOpen
  \bibfield  {author} {\bibinfo {author} {\bibfnamefont {S.}~\bibnamefont
  {Moretti}}\ and\ \bibinfo {author} {\bibfnamefont {K.}~\bibnamefont
  {Yagyu}},\ }\href {\doibase 10.1103/PhysRevD.91.055022} {\bibfield  {journal}
  {\bibinfo  {journal} {Phys. Rev. D}\ }\textbf {\bibinfo {volume} {91}},\
  \bibinfo {pages} {055022} (\bibinfo {year} {2015})},\ \Eprint
  {http://arxiv.org/abs/1501.06544} {arXiv:1501.06544 [hep-ph]} \BibitemShut
  {NoStop}%
\bibitem [{\citenamefont {Moretti}\ \emph {et~al.}(2015)\citenamefont
  {Moretti}, \citenamefont {Rojas},\ and\ \citenamefont
  {Yagyu}}]{Moretti:2015tva}%
  \BibitemOpen
  \bibfield  {author} {\bibinfo {author} {\bibfnamefont {S.}~\bibnamefont
  {Moretti}}, \bibinfo {author} {\bibfnamefont {D.}~\bibnamefont {Rojas}}, \
  and\ \bibinfo {author} {\bibfnamefont {K.}~\bibnamefont {Yagyu}},\ }\href
  {\doibase 10.1007/JHEP08(2015)116} {\bibfield  {journal} {\bibinfo  {journal}
  {JHEP}\ }\textbf {\bibinfo {volume} {08}},\ \bibinfo {pages} {116} (\bibinfo
  {year} {2015})},\ \Eprint {http://arxiv.org/abs/1504.06432} {arXiv:1504.06432
  [hep-ph]} \BibitemShut {NoStop}%
\bibitem [{\citenamefont {Merchand}\ and\ \citenamefont
  {Sher}(2020)}]{Merchand:2019bod}%
  \BibitemOpen
  \bibfield  {author} {\bibinfo {author} {\bibfnamefont {M.}~\bibnamefont
  {Merchand}}\ and\ \bibinfo {author} {\bibfnamefont {M.}~\bibnamefont
  {Sher}},\ }\href {\doibase 10.1007/JHEP03(2020)108} {\bibfield  {journal}
  {\bibinfo  {journal} {JHEP}\ }\textbf {\bibinfo {volume} {03}},\ \bibinfo
  {pages} {108} (\bibinfo {year} {2020})},\ \Eprint
  {http://arxiv.org/abs/1911.06477} {arXiv:1911.06477 [hep-ph]} \BibitemShut
  {NoStop}%
\bibitem [{\citenamefont {Grzadkowski}\ \emph {et~al.}(2009)\citenamefont
  {Grzadkowski}, \citenamefont {Ogreid},\ and\ \citenamefont
  {Osland}}]{Grzadkowski:2009bt}%
  \BibitemOpen
  \bibfield  {author} {\bibinfo {author} {\bibfnamefont {B.}~\bibnamefont
  {Grzadkowski}}, \bibinfo {author} {\bibfnamefont {O.~M.}\ \bibnamefont
  {Ogreid}}, \ and\ \bibinfo {author} {\bibfnamefont {P.}~\bibnamefont
  {Osland}},\ }\href {\doibase 10.1103/PhysRevD.80.055013} {\bibfield
  {journal} {\bibinfo  {journal} {Phys. Rev. D}\ }\textbf {\bibinfo {volume}
  {80}},\ \bibinfo {pages} {055013} (\bibinfo {year} {2009})},\ \Eprint
  {http://arxiv.org/abs/0904.2173} {arXiv:0904.2173 [hep-ph]} \BibitemShut
  {NoStop}%
\bibitem [{\citenamefont {Faro}\ and\ \citenamefont
  {Ivanov}(2019)}]{Faro:2019vcd}%
  \BibitemOpen
  \bibfield  {author} {\bibinfo {author} {\bibfnamefont {F.~S.}\ \bibnamefont
  {Faro}}\ and\ \bibinfo {author} {\bibfnamefont {I.~P.}\ \bibnamefont
  {Ivanov}},\ }\href {\doibase 10.1103/PhysRevD.100.035038} {\bibfield
  {journal} {\bibinfo  {journal} {Phys. Rev. D}\ }\textbf {\bibinfo {volume}
  {100}},\ \bibinfo {pages} {035038} (\bibinfo {year} {2019})},\ \Eprint
  {http://arxiv.org/abs/1907.01963} {arXiv:1907.01963 [hep-ph]} \BibitemShut
  {NoStop}%
\bibitem [{\citenamefont {Lee}\ \emph {et~al.}(1977)\citenamefont {Lee},
  \citenamefont {Quigg},\ and\ \citenamefont {Thacker}}]{Lee:1977eg}%
  \BibitemOpen
  \bibfield  {author} {\bibinfo {author} {\bibfnamefont {B.~W.}\ \bibnamefont
  {Lee}}, \bibinfo {author} {\bibfnamefont {C.}~\bibnamefont {Quigg}}, \ and\
  \bibinfo {author} {\bibfnamefont {H.~B.}\ \bibnamefont {Thacker}},\ }\href
  {\doibase 10.1103/PhysRevD.16.1519} {\bibfield  {journal} {\bibinfo
  {journal} {Phys. Rev. D}\ }\textbf {\bibinfo {volume} {16}},\ \bibinfo
  {pages} {1519} (\bibinfo {year} {1977})}\BibitemShut {NoStop}%
\bibitem [{\citenamefont {Djouadi}(2008{\natexlab{a}})}]{Djouadi:2005gi}%
  \BibitemOpen
  \bibfield  {author} {\bibinfo {author} {\bibfnamefont {A.}~\bibnamefont
  {Djouadi}},\ }\href {\doibase 10.1016/j.physrep.2007.10.004} {\bibfield
  {journal} {\bibinfo  {journal} {Phys. Rept.}\ }\textbf {\bibinfo {volume}
  {457}},\ \bibinfo {pages} {1} (\bibinfo {year} {2008}{\natexlab{a}})},\
  \Eprint {http://arxiv.org/abs/hep-ph/0503172} {arXiv:hep-ph/0503172}
  \BibitemShut {NoStop}%
\bibitem [{\citenamefont {Djouadi}(2008{\natexlab{b}})}]{Djouadi:2005gj}%
  \BibitemOpen
  \bibfield  {author} {\bibinfo {author} {\bibfnamefont {A.}~\bibnamefont
  {Djouadi}},\ }\href {\doibase 10.1016/j.physrep.2007.10.005} {\bibfield
  {journal} {\bibinfo  {journal} {Phys. Rept.}\ }\textbf {\bibinfo {volume}
  {459}},\ \bibinfo {pages} {1} (\bibinfo {year} {2008}{\natexlab{b}})},\
  \Eprint {http://arxiv.org/abs/hep-ph/0503173} {arXiv:hep-ph/0503173}
  \BibitemShut {NoStop}%
\bibitem [{\citenamefont {Aaboud}\ \emph {et~al.}(2018)\citenamefont {Aaboud}
  \emph {et~al.}}]{ATLAS:2018hxb}%
  \BibitemOpen
  \bibfield  {author} {\bibinfo {author} {\bibfnamefont {M.}~\bibnamefont
  {Aaboud}} \emph {et~al.} (\bibinfo {collaboration} {ATLAS}),\ }\href
  {\doibase 10.1103/PhysRevD.98.052005} {\bibfield  {journal} {\bibinfo
  {journal} {Phys. Rev. D}\ }\textbf {\bibinfo {volume} {98}},\ \bibinfo
  {pages} {052005} (\bibinfo {year} {2018})},\ \Eprint
  {http://arxiv.org/abs/1802.04146} {arXiv:1802.04146 [hep-ex]} \BibitemShut
  {NoStop}%
\bibitem [{\citenamefont {Sirunyan}\ \emph
  {et~al.}(2018{\natexlab{b}})\citenamefont {Sirunyan} \emph
  {et~al.}}]{CMS:2018piu}%
  \BibitemOpen
  \bibfield  {author} {\bibinfo {author} {\bibfnamefont {A.~M.}\ \bibnamefont
  {Sirunyan}} \emph {et~al.} (\bibinfo {collaboration} {CMS}),\ }\href
  {\doibase 10.1007/JHEP11(2018)185} {\bibfield  {journal} {\bibinfo  {journal}
  {JHEP}\ }\textbf {\bibinfo {volume} {11}},\ \bibinfo {pages} {185} (\bibinfo
  {year} {2018}{\natexlab{b}})},\ \Eprint {http://arxiv.org/abs/1804.02716}
  {arXiv:1804.02716 [hep-ex]} \BibitemShut {NoStop}%
\bibitem [{\citenamefont {Tumasyan}\ \emph {et~al.}(2022)\citenamefont
  {Tumasyan} \emph {et~al.}}]{CMS:2022qva}%
  \BibitemOpen
  \bibfield  {author} {\bibinfo {author} {\bibfnamefont {A.}~\bibnamefont
  {Tumasyan}} \emph {et~al.} (\bibinfo {collaboration} {CMS}),\ }\href
  {\doibase 10.1103/PhysRevD.105.092007} {\bibfield  {journal} {\bibinfo
  {journal} {Phys. Rev. D}\ }\textbf {\bibinfo {volume} {105}},\ \bibinfo
  {pages} {092007} (\bibinfo {year} {2022})},\ \Eprint
  {http://arxiv.org/abs/2201.11585} {arXiv:2201.11585 [hep-ex]} \BibitemShut
  {NoStop}%
\bibitem [{\citenamefont {Peskin}\ and\ \citenamefont
  {Takeuchi}(1992)}]{Peskin:1991sw}%
  \BibitemOpen
  \bibfield  {author} {\bibinfo {author} {\bibfnamefont {M.~E.}\ \bibnamefont
  {Peskin}}\ and\ \bibinfo {author} {\bibfnamefont {T.}~\bibnamefont
  {Takeuchi}},\ }\href {\doibase 10.1103/PhysRevD.46.381} {\bibfield  {journal}
  {\bibinfo  {journal} {Phys. Rev. D}\ }\textbf {\bibinfo {volume} {46}},\
  \bibinfo {pages} {381} (\bibinfo {year} {1992})}\BibitemShut {NoStop}%
\bibitem [{\citenamefont {Grimus}\ \emph {et~al.}(2008)\citenamefont {Grimus},
  \citenamefont {Lavoura}, \citenamefont {Ogreid},\ and\ \citenamefont
  {Osland}}]{Grimus:2008nb}%
  \BibitemOpen
  \bibfield  {author} {\bibinfo {author} {\bibfnamefont {W.}~\bibnamefont
  {Grimus}}, \bibinfo {author} {\bibfnamefont {L.}~\bibnamefont {Lavoura}},
  \bibinfo {author} {\bibfnamefont {O.~M.}\ \bibnamefont {Ogreid}}, \ and\
  \bibinfo {author} {\bibfnamefont {P.}~\bibnamefont {Osland}},\ }\href
  {\doibase 10.1016/j.nuclphysb.2008.04.019} {\bibfield  {journal} {\bibinfo
  {journal} {Nucl. Phys. B}\ }\textbf {\bibinfo {volume} {801}},\ \bibinfo
  {pages} {81} (\bibinfo {year} {2008})},\ \Eprint
  {http://arxiv.org/abs/0802.4353} {arXiv:0802.4353 [hep-ph]} \BibitemShut
  {NoStop}%
\bibitem [{\citenamefont {Aghanim}\ \emph {et~al.}(2020)\citenamefont {Aghanim}
  \emph {et~al.}}]{Planck:2018vyg}%
  \BibitemOpen
  \bibfield  {author} {\bibinfo {author} {\bibfnamefont {N.}~\bibnamefont
  {Aghanim}} \emph {et~al.} (\bibinfo {collaboration} {Planck}),\ }\href
  {\doibase 10.1051/0004-6361/201833910} {\bibfield  {journal} {\bibinfo
  {journal} {Astron. Astrophys.}\ }\textbf {\bibinfo {volume} {641}},\ \bibinfo
  {pages} {A6} (\bibinfo {year} {2020})},\ \bibinfo {note} {[Erratum:
  Astron.Astrophys. 652, C4 (2021)]},\ \Eprint
  {http://arxiv.org/abs/1807.06209} {arXiv:1807.06209 [astro-ph.CO]}
  \BibitemShut {NoStop}%
\bibitem [{\citenamefont {Aprile}\ \emph {et~al.}(2017)\citenamefont {Aprile}
  \emph {et~al.}}]{XENON:2017vdw}%
  \BibitemOpen
  \bibfield  {author} {\bibinfo {author} {\bibfnamefont {E.}~\bibnamefont
  {Aprile}} \emph {et~al.} (\bibinfo {collaboration} {XENON}),\ }\href
  {\doibase 10.1103/PhysRevLett.119.181301} {\bibfield  {journal} {\bibinfo
  {journal} {Phys. Rev. Lett.}\ }\textbf {\bibinfo {volume} {119}},\ \bibinfo
  {pages} {181301} (\bibinfo {year} {2017})},\ \Eprint
  {http://arxiv.org/abs/1705.06655} {arXiv:1705.06655 [astro-ph.CO]}
  \BibitemShut {NoStop}%
\bibitem [{\citenamefont {Aprile}\ \emph {et~al.}(2018)\citenamefont {Aprile}
  \emph {et~al.}}]{XENON:2018voc}%
  \BibitemOpen
  \bibfield  {author} {\bibinfo {author} {\bibfnamefont {E.}~\bibnamefont
  {Aprile}} \emph {et~al.} (\bibinfo {collaboration} {XENON}),\ }\href
  {\doibase 10.1103/PhysRevLett.121.111302} {\bibfield  {journal} {\bibinfo
  {journal} {Phys. Rev. Lett.}\ }\textbf {\bibinfo {volume} {121}},\ \bibinfo
  {pages} {111302} (\bibinfo {year} {2018})},\ \Eprint
  {http://arxiv.org/abs/1805.12562} {arXiv:1805.12562 [astro-ph.CO]}
  \BibitemShut {NoStop}%
\bibitem [{\citenamefont {Tan}\ \emph {et~al.}(2016)\citenamefont {Tan} \emph
  {et~al.}}]{PandaX-II:2016vec}%
  \BibitemOpen
  \bibfield  {author} {\bibinfo {author} {\bibfnamefont {A.}~\bibnamefont
  {Tan}} \emph {et~al.} (\bibinfo {collaboration} {PandaX-II}),\ }\href
  {\doibase 10.1103/PhysRevLett.117.121303} {\bibfield  {journal} {\bibinfo
  {journal} {Phys. Rev. Lett.}\ }\textbf {\bibinfo {volume} {117}},\ \bibinfo
  {pages} {121303} (\bibinfo {year} {2016})},\ \Eprint
  {http://arxiv.org/abs/1607.07400} {arXiv:1607.07400 [hep-ex]} \BibitemShut
  {NoStop}%
\bibitem [{\citenamefont {Cui}\ \emph {et~al.}(2017)\citenamefont {Cui} \emph
  {et~al.}}]{PandaX-II:2017hlx}%
  \BibitemOpen
  \bibfield  {author} {\bibinfo {author} {\bibfnamefont {X.}~\bibnamefont
  {Cui}} \emph {et~al.} (\bibinfo {collaboration} {PandaX-II}),\ }\href
  {\doibase 10.1103/PhysRevLett.119.181302} {\bibfield  {journal} {\bibinfo
  {journal} {Phys. Rev. Lett.}\ }\textbf {\bibinfo {volume} {119}},\ \bibinfo
  {pages} {181302} (\bibinfo {year} {2017})},\ \Eprint
  {http://arxiv.org/abs/1708.06917} {arXiv:1708.06917 [astro-ph.CO]}
  \BibitemShut {NoStop}%
\bibitem [{\citenamefont {B\'elanger}\ \emph {et~al.}(2015)\citenamefont
  {B\'elanger}, \citenamefont {Boudjema}, \citenamefont {Pukhov},\ and\
  \citenamefont {Semenov}}]{Belanger:2014vza}%
  \BibitemOpen
  \bibfield  {author} {\bibinfo {author} {\bibfnamefont {G.}~\bibnamefont
  {B\'elanger}}, \bibinfo {author} {\bibfnamefont {F.}~\bibnamefont
  {Boudjema}}, \bibinfo {author} {\bibfnamefont {A.}~\bibnamefont {Pukhov}}, \
  and\ \bibinfo {author} {\bibfnamefont {A.}~\bibnamefont {Semenov}},\ }\href
  {\doibase 10.1016/j.cpc.2015.03.003} {\bibfield  {journal} {\bibinfo
  {journal} {Comput. Phys. Commun.}\ }\textbf {\bibinfo {volume} {192}},\
  \bibinfo {pages} {322} (\bibinfo {year} {2015})},\ \Eprint
  {http://arxiv.org/abs/1407.6129} {arXiv:1407.6129 [hep-ph]} \BibitemShut
  {NoStop}%
\bibitem [{\citenamefont {Arina}\ \emph {et~al.}(2009)\citenamefont {Arina},
  \citenamefont {Ling},\ and\ \citenamefont {Tytgat}}]{arina:2009}%
  \BibitemOpen
  \bibfield  {author} {\bibinfo {author} {\bibfnamefont {C.}~\bibnamefont
  {Arina}}, \bibinfo {author} {\bibfnamefont {F.-S.}\ \bibnamefont {Ling}}, \
  and\ \bibinfo {author} {\bibfnamefont {M.~H.}\ \bibnamefont {Tytgat}},\
  }\href {\doibase 10.1088/1475-7516/2009/10/018} {\bibfield  {journal}
  {\bibinfo  {journal} {Journal of Cosmology and Astroparticle Physics}\
  }\textbf {\bibinfo {volume} {2009}},\ \bibinfo {pages} {018} (\bibinfo {year}
  {2009})}\BibitemShut {NoStop}%
\bibitem [{\citenamefont {Alloul}\ \emph {et~al.}(2014)\citenamefont {Alloul},
  \citenamefont {Christensen}, \citenamefont {Degrande}, \citenamefont {Duhr},\
  and\ \citenamefont {Fuks}}]{Alloul:2013bka}%
  \BibitemOpen
  \bibfield  {author} {\bibinfo {author} {\bibfnamefont {A.}~\bibnamefont
  {Alloul}}, \bibinfo {author} {\bibfnamefont {N.~D.}\ \bibnamefont
  {Christensen}}, \bibinfo {author} {\bibfnamefont {C.}~\bibnamefont
  {Degrande}}, \bibinfo {author} {\bibfnamefont {C.}~\bibnamefont {Duhr}}, \
  and\ \bibinfo {author} {\bibfnamefont {B.}~\bibnamefont {Fuks}},\ }\href
  {\doibase 10.1016/j.cpc.2014.04.012} {\bibfield  {journal} {\bibinfo
  {journal} {Comput. Phys. Commun.}\ }\textbf {\bibinfo {volume} {185}},\
  \bibinfo {pages} {2250} (\bibinfo {year} {2014})},\ \Eprint
  {http://arxiv.org/abs/1310.1921} {arXiv:1310.1921 [hep-ph]} \BibitemShut
  {NoStop}%
\bibitem [{\citenamefont {Alwall}\ \emph {et~al.}(2014)\citenamefont {Alwall},
  \citenamefont {Frederix}, \citenamefont {Frixione}, \citenamefont {Hirschi},
  \citenamefont {Maltoni}, \citenamefont {Mattelaer}, \citenamefont {Shao},
  \citenamefont {Stelzer}, \citenamefont {Torrielli},\ and\ \citenamefont
  {Zaro}}]{Alwall:2014hca}%
  \BibitemOpen
  \bibfield  {author} {\bibinfo {author} {\bibfnamefont {J.}~\bibnamefont
  {Alwall}}, \bibinfo {author} {\bibfnamefont {R.}~\bibnamefont {Frederix}},
  \bibinfo {author} {\bibfnamefont {S.}~\bibnamefont {Frixione}}, \bibinfo
  {author} {\bibfnamefont {V.}~\bibnamefont {Hirschi}}, \bibinfo {author}
  {\bibfnamefont {F.}~\bibnamefont {Maltoni}}, \bibinfo {author} {\bibfnamefont
  {O.}~\bibnamefont {Mattelaer}}, \bibinfo {author} {\bibfnamefont {H.~S.}\
  \bibnamefont {Shao}}, \bibinfo {author} {\bibfnamefont {T.}~\bibnamefont
  {Stelzer}}, \bibinfo {author} {\bibfnamefont {P.}~\bibnamefont {Torrielli}},
  \ and\ \bibinfo {author} {\bibfnamefont {M.}~\bibnamefont {Zaro}},\ }\href
  {\doibase 10.1007/JHEP07(2014)079} {\bibfield  {journal} {\bibinfo  {journal}
  {JHEP}\ }\textbf {\bibinfo {volume} {07}},\ \bibinfo {pages} {079} (\bibinfo
  {year} {2014})},\ \Eprint {http://arxiv.org/abs/1405.0301} {arXiv:1405.0301
  [hep-ph]} \BibitemShut {NoStop}%
%%CITATION = ARXIV:1405.0301;%%
\bibitem [{\citenamefont {Sjöstrand}\ \emph {et~al.}(2015)\citenamefont
  {Sjöstrand}, \citenamefont {Ask}, \citenamefont {Christiansen},
  \citenamefont {Corke}, \citenamefont {Desai}, \citenamefont {Ilten},
  \citenamefont {Mrenna}, \citenamefont {Prestel}, \citenamefont {Rasmussen},\
  and\ \citenamefont {Skands}}]{Sjostrand:2014zea}%
  \BibitemOpen
  \bibfield  {author} {\bibinfo {author} {\bibfnamefont {T.}~\bibnamefont
  {Sjöstrand}}, \bibinfo {author} {\bibfnamefont {S.}~\bibnamefont {Ask}},
  \bibinfo {author} {\bibfnamefont {J.~R.}\ \bibnamefont {Christiansen}},
  \bibinfo {author} {\bibfnamefont {R.}~\bibnamefont {Corke}}, \bibinfo
  {author} {\bibfnamefont {N.}~\bibnamefont {Desai}}, \bibinfo {author}
  {\bibfnamefont {P.}~\bibnamefont {Ilten}}, \bibinfo {author} {\bibfnamefont
  {S.}~\bibnamefont {Mrenna}}, \bibinfo {author} {\bibfnamefont
  {S.}~\bibnamefont {Prestel}}, \bibinfo {author} {\bibfnamefont {C.~O.}\
  \bibnamefont {Rasmussen}}, \ and\ \bibinfo {author} {\bibfnamefont {P.~Z.}\
  \bibnamefont {Skands}},\ }\href {\doibase 10.1016/j.cpc.2015.01.024}
  {\bibfield  {journal} {\bibinfo  {journal} {Comput. Phys. Commun.}\ }\textbf
  {\bibinfo {volume} {191}},\ \bibinfo {pages} {159} (\bibinfo {year}
  {2015})},\ \Eprint {http://arxiv.org/abs/1410.3012} {arXiv:1410.3012
  [hep-ph]} \BibitemShut {NoStop}%
%%CITATION = ARXIV:1410.3012;%%
\bibitem [{\citenamefont {de~Favereau}\ \emph {et~al.}(2014)\citenamefont
  {de~Favereau}, \citenamefont {Delaere}, \citenamefont {Demin}, \citenamefont
  {Giammanco}, \citenamefont {Lemaître}, \citenamefont {Mertens},\ and\
  \citenamefont {Selvaggi}}]{deFavereau:2013fsa}%
  \BibitemOpen
  \bibfield  {author} {\bibinfo {author} {\bibfnamefont {J.}~\bibnamefont
  {de~Favereau}}, \bibinfo {author} {\bibfnamefont {C.}~\bibnamefont
  {Delaere}}, \bibinfo {author} {\bibfnamefont {P.}~\bibnamefont {Demin}},
  \bibinfo {author} {\bibfnamefont {A.}~\bibnamefont {Giammanco}}, \bibinfo
  {author} {\bibfnamefont {V.}~\bibnamefont {Lemaître}}, \bibinfo {author}
  {\bibfnamefont {A.}~\bibnamefont {Mertens}}, \ and\ \bibinfo {author}
  {\bibfnamefont {M.}~\bibnamefont {Selvaggi}} (\bibinfo {collaboration}
  {DELPHES 3}),\ }\href {\doibase 10.1007/JHEP02(2014)057} {\bibfield
  {journal} {\bibinfo  {journal} {JHEP}\ }\textbf {\bibinfo {volume} {02}},\
  \bibinfo {pages} {057} (\bibinfo {year} {2014})},\ \Eprint
  {http://arxiv.org/abs/1307.6346} {arXiv:1307.6346 [hep-ex]} \BibitemShut
  {NoStop}%
%%CITATION = ARXIV:1307.6346;%%
\bibitem [{\citenamefont {Cowan}\ \emph {et~al.}(2011)\citenamefont {Cowan},
  \citenamefont {Cranmer}, \citenamefont {Gross},\ and\ \citenamefont
  {Vitells}}]{Cowan:2010js}%
  \BibitemOpen
  \bibfield  {author} {\bibinfo {author} {\bibfnamefont {G.}~\bibnamefont
  {Cowan}}, \bibinfo {author} {\bibfnamefont {K.}~\bibnamefont {Cranmer}},
  \bibinfo {author} {\bibfnamefont {E.}~\bibnamefont {Gross}}, \ and\ \bibinfo
  {author} {\bibfnamefont {O.}~\bibnamefont {Vitells}},\ }\href {\doibase
  10.1140/epjc/s10052-011-1554-0} {\bibfield  {journal} {\bibinfo  {journal}
  {Eur. Phys. J. C}\ }\textbf {\bibinfo {volume} {71}},\ \bibinfo {pages}
  {1554} (\bibinfo {year} {2011})},\ \bibinfo {note} {[Erratum: Eur.Phys.J.C
  73, 2501 (2013)]},\ \Eprint {http://arxiv.org/abs/1007.1727} {arXiv:1007.1727
  [physics.data-an]} \BibitemShut {NoStop}%
\bibitem [{\citenamefont {{Hoecker}}\ \emph {et~al.}(2007)\citenamefont
  {{Hoecker}}, \citenamefont {{Speckmayer}}, \citenamefont {{Stelzer}},
  \citenamefont {{Therhaag}}, \citenamefont {{von Toerne}}, \citenamefont
  {{Voss}}, \citenamefont {{Backes}}, \citenamefont {{Carli}}, \citenamefont
  {{Cohen}}, \citenamefont {{Christov}}, \citenamefont {{Dannheim}},
  \citenamefont {{Danielowski}}, \citenamefont {{Henrot-Versille}},
  \citenamefont {{Jachowski}}, \citenamefont {{Kraszewski}}, \citenamefont
  {{Krasznahorkay}}, \citenamefont {{Kruk}}, \citenamefont {{Mahalalel}},
  \citenamefont {{Ospanov}}, \citenamefont {{Prudent}}, \citenamefont
  {{Robert}}, \citenamefont {{Schouten}}, \citenamefont {{Tegenfeldt}},
  \citenamefont {{Voigt}}, \citenamefont {{Voss}}, \citenamefont {{Wolter}},\
  and\ \citenamefont {{Zemla}}}]{2007physics3039H}%
  \BibitemOpen
  \bibfield  {author} {\bibinfo {author} {\bibfnamefont {A.}~\bibnamefont
  {{Hoecker}}}, \bibinfo {author} {\bibfnamefont {P.}~\bibnamefont
  {{Speckmayer}}}, \bibinfo {author} {\bibfnamefont {J.}~\bibnamefont
  {{Stelzer}}}, \bibinfo {author} {\bibfnamefont {J.}~\bibnamefont
  {{Therhaag}}}, \bibinfo {author} {\bibfnamefont {E.}~\bibnamefont {{von
  Toerne}}}, \bibinfo {author} {\bibfnamefont {H.}~\bibnamefont {{Voss}}},
  \bibinfo {author} {\bibfnamefont {M.}~\bibnamefont {{Backes}}}, \bibinfo
  {author} {\bibfnamefont {T.}~\bibnamefont {{Carli}}}, \bibinfo {author}
  {\bibfnamefont {O.}~\bibnamefont {{Cohen}}}, \bibinfo {author} {\bibfnamefont
  {A.}~\bibnamefont {{Christov}}}, \bibinfo {author} {\bibfnamefont
  {D.}~\bibnamefont {{Dannheim}}}, \bibinfo {author} {\bibfnamefont
  {K.}~\bibnamefont {{Danielowski}}}, \bibinfo {author} {\bibfnamefont
  {S.}~\bibnamefont {{Henrot-Versille}}}, \bibinfo {author} {\bibfnamefont
  {M.}~\bibnamefont {{Jachowski}}}, \bibinfo {author} {\bibfnamefont
  {K.}~\bibnamefont {{Kraszewski}}}, \bibinfo {author} {\bibfnamefont
  {A.}~\bibnamefont {{Krasznahorkay}}, \bibfnamefont {Jr.}}, \bibinfo {author}
  {\bibfnamefont {M.}~\bibnamefont {{Kruk}}}, \bibinfo {author} {\bibfnamefont
  {Y.}~\bibnamefont {{Mahalalel}}}, \bibinfo {author} {\bibfnamefont
  {R.}~\bibnamefont {{Ospanov}}}, \bibinfo {author} {\bibfnamefont
  {X.}~\bibnamefont {{Prudent}}}, \bibinfo {author} {\bibfnamefont
  {A.}~\bibnamefont {{Robert}}}, \bibinfo {author} {\bibfnamefont
  {D.}~\bibnamefont {{Schouten}}}, \bibinfo {author} {\bibfnamefont
  {F.}~\bibnamefont {{Tegenfeldt}}}, \bibinfo {author} {\bibfnamefont
  {A.}~\bibnamefont {{Voigt}}}, \bibinfo {author} {\bibfnamefont
  {K.}~\bibnamefont {{Voss}}}, \bibinfo {author} {\bibfnamefont
  {M.}~\bibnamefont {{Wolter}}}, \ and\ \bibinfo {author} {\bibfnamefont
  {A.}~\bibnamefont {{Zemla}}},\ }\href@noop {} {\bibfield  {journal} {\bibinfo
   {journal} {ArXiv Physics e-prints}\ } (\bibinfo {year} {2007})},\ \Eprint
  {http://arxiv.org/abs/physics/0703039} {physics/0703039} \BibitemShut
  {NoStop}%
\end{thebibliography}%
 %%%%%%%%%%%%%%%%%%%%%%%%%%%%%%%%%%%%%%%%%%%%%%%%%%%%%%%%%%%%%%%%%%%%%
\end{document}